%% file: 0.paper.tex
\pdfoutput=1

\documentclass[sigconf]{acmart}

\usepackage{subfiles} 

\copyrightyear{2021}
\acmYear{2021}
\setcopyright{acmcopyright}\acmConference[FAccT '21]{Conference on Fairness, Accountability, and Transparency}{March 3--10, 2021}{Virtual Event}
\acmBooktitle{Conference on Fairness, Accountability, and Transparency (FAccT '21), March 3--10, 2021, Virtual Event, Canada}
\acmPrice{15.00}
\acmDOI{10.1145/3442188.3445885}
\acmISBN{978-1-4503-8309-7/21/03}

\begin{document}

\title{Data Leverage: A Framework for Empowering the Public in its Relationship with Technology Companies}


\author{Nicholas Vincent}
\affiliation{\institution{Northwestern University}}
\email{nickvincent@u.northwestern.edu}

\author{Hanlin Li}
\affiliation{\institution{Northwestern University}}
\email{lihanlin@u.northwestern.edu}

\author{Nicole Tilly}
\affiliation{\institution{Northwestern University}}
\email{nicoletilly2023@u.northwestern.edu}

\author{Stevie Chancellor}
\affiliation{\institution{University of Minnesota}}
\email{steviec@umn.edu}
\authornote{Chancellor completed much of this work while at Northwestern University.}
\author{Brent Hecht}
\affiliation{\institution{Northwestern University}}
\email{bhecht@northwestern.edu}



\subfile{1.abstract}

\ccsdesc[500]{Human-centered computing~Collaborative and social computing theory, concepts and paradigms}


\keywords{data leverage, data strikes, data poisoning, conscious data contribution}

\maketitle


\subfile{2.intro}

\subfile{3.rw}

\subfile{4.framework}

\subfile{5.assess}

\subfile{6.discuss}



\bibliographystyle{ACM-Reference-Format}
\bibliography{manual}

\end{document}

%% file: 1.abstract.tex
\begin{abstract}
Many powerful computing technologies rely on implicit and explicit data contributions from the public. This dependency suggests a potential source of leverage for the public in its relationship with technology companies: by reducing, stopping, redirecting, or otherwise manipulating data contributions, the public can reduce the effectiveness of many lucrative technologies. In this paper, we synthesize emerging research that seeks to better understand and help people action this \textit{data leverage}. Drawing on prior work in areas including machine learning, human-computer interaction, and fairness and accountability in computing, we present a framework for understanding data leverage that highlights new opportunities to change technology company behavior related to privacy, economic inequality, content moderation and other areas of societal concern. Our framework also points towards ways that policymakers can bolster data leverage as a means of changing the balance of power between the public and tech companies.
\end{abstract}

%% file: 2.intro.tex
\section{Introduction}
In August 2020, the most valuable five technology companies had a total market cap of US\$7 trillion \cite{PaulR.LaMonica2020Oct}. This valuation is driven in part by large models that use data generated by the public to recommend content, rank search results, and provide many other services \cite{brynjolfsson2014second,posner_radical_2018,zhou_impact_2010, mcmahon_substantial_2017}. More generally, lucrative technologies used by many companies rely on data generated by large groups of people to fulfill critical customer needs \cite{brynjolfsson2014second,arrieta_ibarra_should_2018,posner_radical_2018, vincent2018examining} and drive decision-making~\cite{brynjolfsson_rapid_2016}. 

The reliance of powerful technologies (and thus powerful companies) on ``data labor'' \cite{arrieta_ibarra_should_2018,posner_radical_2018,vincent_data_2019} by the general public presents an enormous opportunity for the public to gain more power in its relationship with tech companies. People perform data labor when they engage in the multitude of interactions with technology that generate data for firms (e.g. liking, clicking, rating, posting). By leveraging tech companies' reliance on their data labor, the public could demand changes on pressing issues \cite{hecht_its_2018,kulynych_pots_2020},
such as diminished privacy \cite{pmlr-v81-ekstrand18a,brunton_obfuscation_2015}, the reinforcement of problematic societal biases by AI systems ~\cite{abebe_roles_2020,eubanks2018automating,noble2018algorithms,kulynych_pots_2020,albert2020politics},
eroded labor rights~\cite{hara_data-driven_2018,Paul2020Apr},
environmental harms~\cite{schwartz2019green},
content moderation challenges~\cite{gillespie2018custodians},
and the current imbalance in how profits from data-driven technologies are distributed between tech operators and data contributors \cite{posner_radical_2018,vincent_mapping_2019,brynjolfsson2014second}. Armed with the knowledge of the importance of data contributions and the tools to action this knowledge, the public could potentially interfere with recommender systems, search engines, image classifiers, and other technologies until tech companies made changes related to these issues.

To capture the power inherent in the public's data labor, this paper introduces the concept of ``data leverage'' and discusses how the concept can be made operational.
Simply put, data leverage refers to influence that members of the public have over tech companies because important computing technologies rely on the public's data contributions. Data leverage catalyzes power achieved by \textit{harming} data-dependent technologies
as well as power achieved by \textit{improving} alternative data-dependent technologies and thereby creating increased competition \cite{vincent_can_2020}.
The concept of data leverage highlights an emergent theme in the FAccT community and related areas, including human-computer interaction (HCI), social computing, society and technology studies (STS), machine learning (ML), and particularly ML research that seeks to advance fairness, justice, and a human-centered perspective (e.g. \cite{abebe_roles_2020,green_fair_2018,binns2018fairness,chancellor2019human}). This paper shows that this interdisciplinary lens can provide a structure for understanding and actioning an almost entirely untapped source of power that can advance a wide variety of pro-social goals. Our data leverage framework also highlights opportunities for future research and policy interventions that empower the public in its relationship with technology companies.


The contributions of this work are to (1) define data leverage, (2) provide a framework of potential ``data levers'', grounded in prior work that has advanced our understanding of these levers, (3) outline an initial assessment of strengths and weaknesses of each data lever in the public's ``tool belt'', and (4) highlight how data leverage provides important opportunities for research and new policy. Critically, research and policy can amplify data leverage and, conversely, using data leverage as a lens can raise the stakes for related research areas and policy discussions. We pay particular attention to factors that might facilitate the use of data leverage (e.g. policy interventions) or that block groups from exerting data leverage (drawing on the literature on "non-use" of technology) \cite{satchell_beyond_2009,wyatt2003non, baumer2013limiting,baumer2015missing,baumer2018socioeconomic,baumer2019all,lampe_users_2013}.

\subsection{Background and Definitions}
Before continuing, we first present formal definitions of data leverage and supporting concepts. Note that while aiming to be comprehensive, these are working definitions. Data leverage is an emerging topic in a rapidly moving field, and we aim to advance and open the discussion around data leverage, not conclude it.

\begin{itemize}


\item \textit{Data leverage}: The power derived from computing technologies' dependence on human-generated data. Data leverage 
is exerted when a
group influences an organization by threatening to engage in or directly engaging in data-related actions that harm that organization's technologies or help its competitors' technologies. 

\item \textit{Data levers}: The specific types of actions that individuals or groups engage in  to exert data leverage. For instance, ``data strikes'' \cite{vincent_data_2019} are one of the data levers we discuss below and they operate by cutting off the flow of data to tech companies. 

\end{itemize}

%% file: 3.rw.tex
\section{Related Work}
In this section, we situate data leverage in relation to the FAccT domain, and then discuss four additional areas that contribute to the idea of data leverage.

\subsection{Data Leverage and FAccT Research}

Data leverage emerges in part from work in the broader FAccT community that has demonstrated the limitations of purely technical approaches to advancing fairness and justice in computing systems \cite{abebe_roles_2020,binns2018fairness,eubanks2018automating,gebru2019oxford,chancellor2019human}. This large literature emphasizes the critical roles played by the societal context around computing systems, and has demonstrated that sociotechnical approaches are often much more powerful than purely technical approaches. Data leverage can in many ways be understood as a framework that helps us better understand data-driven technologies through a sociotechnical lens and use that lens to take action to achieve pro-social outcomes.

Data leverage is more specifically informed by \citeauthor{kulynych_pots_2020}'s work that proposed ``Protective Optimization Technologies'' (POTs) as a way to address the negative impacts of algorithmic systems and give agency to those impacted~\cite{kulynych_pots_2020}. POTs allow people to contest or subvert optimization technologies, perhaps adopting techniques from data poisoning (which we further address below)~\cite{kulynych_pots_2020,troncoso_keynote_2019}. Data leverage and POTs are synergistic concepts, and many POTs enable people to exert data leverage.

\subsection{Data as Labor}
Data leverage is heavily informed by work that views data generated by people using computing systems as a type of labor. Building on \citeauthor{posner_radical_2018}~\cite{posner_radical_2018}, \citeauthor{arrieta_ibarra_should_2018} argue that data should be considered as labor, not ``exhaust'' emitted in the process of using technology, and as such, should be subject to some kind of remuneration \cite{arrieta_ibarra_should_2018}. 
The relationship between the data-generating public and the companies that benefit from data is very asymmetric. Not only do people have very little knowledge of --- let alone agency over --- how data they contribute is used, but the economic winnings from powerful data-dependent technologies are reaped entirely by tech companies \cite{posner_radical_2018}. To mitigate this inequality, \citeauthor{posner_radical_2018} called for the formation of ``data unions'', which allow data laborers to collectively negotiate with technology companies~\cite{posner_radical_2018}.

The discussion around data labor has inspired work that aims to measure the economic value of data
\cite{vincent2018examining,vincent2019measuring, koh2019accuracy,mcmahon_substantial_2017}. One approach has been to look at the relationship between Wikipedia --- the product of data contributions from the public --- and real-world economic outcomes such as tourism and investment
\cite{hinnosaar2019wikipedia,Xu2013Dec}. Building on the data as labor concept, Vincent and colleagues have investigated how people might withhold or redirect their data labor to force a data-dependent organization to change its practices~\cite{vincent_can_2020,vincent_data_2019}.

Scholars working on data feminism --- an intersectional feminism-informed lens for data science  --- have called for more efforts to make the labor of data science visible, including the labor of data generation \cite{d2020data}. These scholars argue that the invisible labor of data science, much like housework, has been hidden from public view and therefore undervalued \cite{d2020data}, and that researchers can begin to shine a light on this labor by studying and highlighting the processes of data creation (e.g. \cite{crawford2018anatomy}). In this way, data feminism is very aligned with the ideas of data leverage; both aim to measure and make people aware of the value of previously invisible labor and ultimately reshape power imbalances.




\subsection{Data Leverage and Technology Use/Non-Use}
The data leverage concept is also informed by work from HCI and STS on technology ``use'', ``non-use'', and the spectrum of behaviors in between. 



Work from \citeauthor{selwyn2003apart} and \citeauthor{wyatt2003non} called attention to the need to understand people who do not use new technologies~\cite{selwyn2003apart,wyatt2003non}.
Most relevant to data leverage, \citeauthor{selwyn2003apart} documented that people engage in ideological refusal to use certain technology ``despite being able to do so in practice''. Further calls to study non-use in HCI and STS have been amplified in the years since \cite{satchell_beyond_2009, baumer2014refusing}.

Use and non-use exist on a spectrum \cite{wyatt2003non,baumer2019all}. People face many social and technical decisions in terms of when they will use, and stop using, a particular technology, and these decisions lead to many different forms of use and non-use \cite{brubaker2016departing, schoenebeck_giving_2014, baumer2013limiting}. Recently, \citeauthor{saxena2020methods} reviewed the methods for creating typologies to describe the many forms of use and non-use \cite{saxena2020methods}. 

Many factors motivate non-use, such as exclusion \cite{wyatt2003non}, social capital \cite{lampe_users_2013}, and socioeconomic factors \cite{baumer2018socioeconomic,baumer2019all}. Anyone seeking to use data leverage to empower the public must contend with these factors. Attempts to support data leverage could exclude or disproportionately benefit certain groups following existing patterns in how technology excludes and benefits these groups.

One common theme in the non-use literature is that it is not easy for people to refrain from use when it comes to products that have some benefit in their life, even if the benefit(s) come with a host of long-term drawbacks. People often speak of their technology use as a type of addiction, using terms like `relapsing'' and ``withdrawing''  \cite{baumer2015missing, baumer2013limiting}. Challenges also emerge related to the public presentation role of social media profiles ~\cite{lampe_use_perception}. Even if people stop using a technology, they may not necessarily delete their data. In studying individuals who left Grindr, a dating app, \citeauthor{brubaker2016departing} found that ``even among those who deleted the app, only a minority tried to close their accounts or remove personal data...[putting them] in a paradoxical position of thinking they have left while their profile --- or data --- continues on'' \cite{brubaker2016departing}.

The non-use literature also indicates that people engage in protest-related use and non-use behaviors for reasons relating to privacy, data practices, perceived addiction, and other issues ~\cite{baumer2013limiting,baumer2015missing,li_how_2019,stieger_who_2013}. Anyone engaging in such behaviors is a potential participant in data leverage campaigns. \citeauthor{casemajor_non-participation_2015} and \citeauthor{portwood2013media} argue separately that non-participation in digital media can be an explicitly political action\cite{casemajor_non-participation_2015, portwood2013media}. \citeauthor{li_how_2019} conducted a survey to better understand ``protest users'', or people who stop or change their use of tech to protest tech companies\cite{li_how_2019}. The results suggested that there is a large number of people interested in protest use: half of respondents were interested in becoming protest users of a company, and 30\% were already engaged in protest use.

An important related lens is that of ``refusal''. Focusing on bioethics, Benjamin makes the case that broad support of ``informed refusal'' provides a means of developing a justice-oriented paradigm of science and technology \cite{benjamin_informed_2016}. In practice, people who engage in informed refusal are engaging in a political form of non-use, and thereby data leverage. Building on Benjamin's work, \citeauthor{cifor2019feminist} and \citeauthor{garcia2020no} describe how the notion of ``critical refusal'' informed by feminist scholars can be used improve practices around data \cite{garcia2020no,cifor2019feminist}.

\subsection{Data Leverage and ML Research} Understanding the full potential of data leverage requires deep engagement with machine learning literature. Two relevant areas of ML research are those that answer questions around (1) the effectiveness of adversarial attacks on data-dependent systems and (2) the relationship between a system's performance and changes to underlying data.

There is a large literature that considers the case of adversaries attempting to attack ML systems (e.g. \cite{barreno2006can,pitropakis_taxonomy_2019,biggio2012poisoning,shan2020fawkes,shafahi2018poison,steinhardt2017certified,lam2004shilling,lee_shilling_2012,gunes_shilling_2014,chirita_preventing_2005,mobasher2005effective,Fang2020Apr,li2016data}).
In early work on adversarial ML, \citeauthor{barreno2006can} developed a taxonomy of attacks on ML systems~\cite{barreno2006can}. They focused in particular on attacks in which an adversary ``mis-trains'' a system, which is called \textit{data poisoning}. Data poisoning attacks against many types of ML systems have been studied in detail \cite{barreno2006can,pitropakis_taxonomy_2019,biggio2012poisoning,shan2020fawkes,shafahi2018poison,steinhardt2017certified}.
A type of data poisoning attack that is particularly relevant to the work in this paper is the ``shilling'' attack, which involves ``lying'' to a recommender system so that a system recommends certain products favored by the attacker \cite{lam2004shilling}. Accordingly, much work has been done on counteracting shilling (e.g. \cite{lee_shilling_2012,gunes_shilling_2014,chirita_preventing_2005,mobasher2005effective}), which may be of concern to groups who want to use shilling-style data poisoning attacks to exert data leverage as we describe below. Researchers have also explored advanced ``data poisoning'' techniques that use sophisticated methods to optimally harm ML systems \cite{Fang2020Apr,li2016data}, which can be much more effective than unsophisticated attacks (e.g. providing random or average ratings to many items \cite{lam2004shilling}). 

Data leverage raises the stakes of the already high-stakes adversarial ML domain. This paper highlights how adversarial techniques, such as data poisoning, are not just relevant to issues of security and privacy, but also to the power dynamics between users and tech companies. While some recent work in adversarial ML has taken a political lens and highlighted real world examples of how adversarial ML can create socially desirable outcomes \cite{albert2020politics}, most of the literature takes a strictly security-oriented lens. 

The literature on the relationship between the amount of training data a model has access to and model performance is also highly relevant to data leverage. Many authors have found diminishing returns of additional data across many contexts and algorithms (e.g. \cite{cho2015much,dietterich_machine_1995,figueroa_predicting_2012,hestness2017deep}), and some have studied techniques to address diminishing returns \cite{bloodgood_bucking_2010}. These findings are informative as to how effective data leverage can be.

\subsection{Data Leverage and Data Activism}
This paper builds on the literature that explores how the public can change practices of the technology industry. Data activism is a relatively new form of civic participation in response to tech companies' pervasive role in public life \cite{Baack}.

Currently, data activism encompasses practices that affect technology design, development, and deployment \cite{milan2016alternative}. Data leverage can be seen as a subset of data activism with a specific focus on empowering the public to influence the performance of data-dependent technologies. \citeauthor{milan2016alternative} provided a typology of data activism that further illustrated the specialized activities in this space --- proactive and reactive data activism \cite{milan2016alternative}. Proactive data activism refers to activists directly influencing software development or databases through open source projects or collaborating with institutions. A particularly relevant data activism initiative is the open data movement, which aims to democratize information that is currently only accessible to the state or businesses \cite{Gurstein}. For example, \citeauthor{Baack} studied an open data project in Finland and highlighted the intermediary role of data activists between the public and operators of data-dependent technologies~\cite{Baack}. On the other hand, reactive data activism entails activists acting against data-collecting entities through adversarial behaviors such as employing encryption. Data leverage includes both types of data activism.

Equipped with the knowledge and expertise to understand data's role in computing, researchers can provide the public with valuable information to identify and employ effective data leverage practices. Work on data activism has unveiled a rich space to improve data practices \cite{couldry2014big}. In particular, \citeauthor{lehtiniemi2018social} called for ``linking knowledge production to data activism practice'' to gain a comprehensive understanding of data's role in the public sphere \cite{lehtiniemi2018social}.

%% file: 4.framework.tex
\section{Data Leverage Framework}
In this section, we describe our framework for data leverage in detail . The framework --- and this section --- is organized around the three data levers we identified. For each lever, we first define the lever and any variants, and do so grounded in past work viewed through our data leverage lens. We then provide practical examples of each data lever and describe the likely factors that will govern the effectiveness of the  lever. Table 1 lists the data levers, their definitions, and several examples of each.



\begin{table*}
  \caption{The three data levers in our framework, short definitions for each, and several examples of each.}
  \label{tab:commands}
  \begin{tabular} {p{4cm}p{4.5cm}p{7cm}}
    \toprule
    \textbf{Data Lever Name}&
    \textbf{Short Definition}&
    \textbf{Examples}\\
    
    \midrule
    Data Strike&
    withholding or deleting data&
    leaving a platform, installing privacy tools \\
    
    Data Poisoning&
    contributing harmful data&
    inputting fake data in user profile, clicking randomly, manipulating images\\
    
    Conscious Data Contribution&
    contributing data to a competitor&
    switching to a new search engine,
    transferring photos to a new platform\\
    \bottomrule
  \end{tabular}
\end{table*}

\subsection{Data Strikes}

The first of the data levers we will consider are data strikes. Data strikes involve a person withholding or deleting data to reduce the amount of data an organization has available to train and operate data-dependent technologies. Although the term data strike is relatively new, the concept builds on the well-studied practices of stopping or changing technology use as a form of protest, as discussed in Related Work. For instance, groups have participated in prominent boycotts against companies like Facebook and Uber \cite{Semuels2017Feb,greenfield_naacp_2018}. In another example, people use ad blocking software to deprive companies of data about the success of their ad placements \cite{budak_understanding_2016}.

\subsubsection{Data Strike Variants} 

The most basic form of a data strike is a \textit{withholding-based data strike}. In some cases, users can withhold data by reducing or stopping their technology use, or by continuing to use a technology with privacy-protection tools (e.g. tracking blockers \cite{mathur2018characterizing}). In jurisdictions that allow people to delete their past data (using laws like the General Data Protection Regulation (GDPR) and California Consumer Privacy Act (CCPA) \cite{Wakabayashi2018Jun,satariano_what_2020}), users can also engage in \textit{deletion-based data strikes}. The effectiveness of such strikes will depend on how well regulations can force companies to regularly retrain or delete their models (so as to remove weights learned using now-deleted data). There is some precedent from the U.S. that model deletion can be enforced: in 2021, the Federal Trade Commission forced a company to delete both customer photos and the facial recognition models trained on the photos \cite{lyons_ftc_2021}.

Data strikes can be further categorized based on their coordination requirements. Data strikes (and ther other data levers we will describe below) are likely possible without serious coordination, given the success of hashtag activism \cite{jackson2020hashtagactivism} and other forms of online collective action that operate without central leadership~ \cite{margetts2015political}. For instance, people wanting to start an informally-organized data strike might simply make a call for others to delete as much data as they are willing. However, ``targeted'' \cite{barreno2006can} data strikes have the potential for a group of data strikers to achieve disproportionate impact \cite{vincent_data_2019}. Following \citeauthor{barreno2006can}'s definition of targeted attacks on ML systems, a targeted data strike might encourage participants to delete specific data points or recruit particularly valuable participants. For example, data strikers could try to reduce performance for a specific genre of movie recommendations, while leaving performance for other genres untouched \cite{vincent_data_2019}. Leaders might also recruit specific users to join their data strike -- power users have disproportionate influence on systems \cite{wilson_evil_2014,wilson_when_2013,eskandanian_power_2019} and withholding or deleting their data may be more impactful.

\subsubsection{What Do Data Strikes Look Like in Practice?}
To understand what data strikes will look like, we can gain insight from the non-use literature described above. An individual that chooses to use a platform less frequently or avoid a feature of that platform reduces the amount of data they help to generate. In this way, a person's choices about use and non-use affect how much data that person generates. Research in the use and non-use domain has provided empirical examples of what could be conceptualized as data strikes against Facebook and Twitter~\cite{baumer2013limiting,baumer2015missing,baumer2018socioeconomic,baumer2019all,schoenebeck_giving_2014,portwood2013media}. 

Privacy and surveillance research also lends itself to uncovering privacy-focused behaviors that can be seen as data strikes. One prominent example is that many people use anti-tracking browser extensions that limit the amount of data online trackers collect \cite{li_how_2019, mathur2018characterizing,budak_understanding_2016}. Studies on algorithm transparency also provide evidence suggesting that people engage with data strike-like behaviors because of dissatisfaction with algorithmic system opacity, such as ceasing producing reviews for review platforms \cite{eslami2017careful, eslami2019user}. Additionally, research on online communities presented case studies of both Reddit moderators and community members striking by disabling and leaving their communities \cite{matias2016going, newell2016user}. 

\subsubsection{How Can Data Strikes Be Effective?}
A data strike can be evaluated based on the importance of the data that ``goes missing'' in terms of how that data affects relevant data-dependent systems. Said another way, does the missing data noticeably degrade a system's performance, move a classifier's decision boundary (or hyperplane, etc.) in a meaningful way, or otherwise change outputs?

To understand the effectiveness of data strikes, researchers and strike leaders might look to research on data scaling and learning curves, which describes the relationship between ML performance and the amount of data available for training (e.g. \cite{cho2015much,dietterich_machine_1995,figueroa_predicting_2012,hestness2017deep}). Findings from this literature could be used to predict the effectiveness of a strike, as in prior work which explicitly simulated data strikes ~\cite{vincent_data_2019,vincent_can_2020}. 
If researchers have shown a model needs a certain number of observations in its training set to be effective (e.g. \cite{cho2015much}), data strike organizers could use that research to guide their strike, for instance by setting a goal for participant recruitment.

In summary, data strikes are a data lever available to anyone who can withhold or delete their data. While a new concept, research in HCI, privacy, machine learning, and related fields can help us to understand what data strikes will look like and how effective they might be.

\subsection{Data Poisoning}
A data poisoning attack is an adversarial attack that inserts inaccurate or harmful training data into a data-dependent technology, thereby encouraging the model to perform poorly \cite{barreno2006can}. While data strikes harm performance by reducing the amount of available data, data poisoning harms performance by providing a technology with data that was created with the intention of thwarting the technology. A relatively accessible way that users can engage in data poisoning is simply by leveraging standard technology features in a deceptive manner. For instance, someone who dislikes pop music might use an online music platform to play a playlist of pop music when they step away from their device with the intention of ``tricking'' a recommender system into using their data to recommend pop music to similar pop-hating users. Other straightforward examples include the coordinated effort to create sexually explicit Google search results for former U.S. Senator Rick Santorum's name ~\cite{gillespie2017algorithmically} and coordinated campaigns to use fake reviews to promote certain products \cite{lam2004shilling}. As we will describe below, very sophisticated variants of data poisoning that draw on state-of-the-art machine learning research are also possible. 

\subsubsection{Data Poisoning Variants}

Data poisoning is familiar to the ML community through adversarial ML (see e.g \cite{barreno2006can,pitropakis_taxonomy_2019,biggio2012poisoning,shan2020fawkes,shafahi2018poison,steinhardt2017certified}) and obfuscation (see e.g. \cite{brunton_obfuscation_2015, howe2017engineering}). This means data poisoning organizers can benefit from the knowledge produced through this body of research.

There are many ways an individual alone can engage in data poisoning. 
The techniques for obfuscation described by \citeauthor{brunton_obfuscation_2015} are accessible means of data poisoning for individuals. For instance, users might trade accounts (drawing on \citeauthor{brunton_obfuscation_2015}) or fill in parts of their profile with fake information \cite{cho2011unsell}. As another example, past work has studied attacks that involve following certain Twitter users to throw off Twitter's profiling \cite{nechaev_concealing_2017}. 
These approaches are generally available to an individual acting alone.

The distinction between coordinated data poisoning attacks and uncoordinated attacks is important. Typically, adversarial ML papers frame data poisoning as a contest between a single attacker (which could be an organization) and a defender/victim. In a coordinated data poisoning attack, however, the attacker is an organized collective. 

To execute a coordinated data poisoning attack, it will be necessary to find the appropriate technique for a particular technology. Organizers can look to taxonomies in the adversarial ML literature to see what knowledge an attacker requires and what specific systems are vulnerable to attacks \cite{barreno2006can,pitropakis_taxonomy_2019}. 

Shilling attacks are a data poisoning variant that focuses on manipulating specific system outcomes rather than general performance degradation \cite{lam2004shilling,lee_shilling_2012,gunes_shilling_2014}. Unlike other poisoning attacks, this type of data leverage manipulates a system to favorably recommend a product that may not actually be high in quality or popularity, i.e. putting ``lipstick on a pig''. Shilling can be defended against with systems that identify and remove fraudulent or false reviews \cite{ott2012estimating, li2014towards}, but these systems themselves may be vulnerable to data poisoning and data strikes.

As with other forms of data leverage, data poisoning applies more generally to any data-dependent technology, not just to ML systems. For instance, \citeauthor{tahmasebian2020crowdsourcing} provide a taxonomy of data poisoning attacks against crowdsourcing-based ``truth inference'' systems \cite{tahmasebian2020crowdsourcing}, e.g. a system that aims to use crowdsourced data to ascertain the true number of cars on a road. Generally, any system that makes or uses estimates about a population can be compromised by sampling poisoned data.

\subsubsection{What Does Data Poisoning Look Like in Practice?}
Almost any data-driven technology is vulnerable to deceptive interactions from users, and there are numerous ways to engage in data poisoning in practice. In the wild, there are a wide range of behaviors that constitute  data poisoning attacks. Examples include Uber and Lyft drivers providing false information about their availability \cite{lee2015working} and internet-browsing user using software to automatically click ads \cite{howe2017engineering}. 

The most accessible form of data poisoning involves a person using technology in a deceptive manner, e.g. by lying about their personal attributes, watching videos they dislike, or searching for content they are not interested in. They might even use deception-support tools like the location-spoofing software conceptualized by \citeauthor{10.1145/2858036.2858060} to engage in ``computationally-mediated pro-social deception''~\cite{10.1145/2858036.2858060}.

By combining findings and tools from HCI and ML, more complex forms of data poisoning may be possible. Users might employ tools like browser extensions (following \citeauthor{li_out_2018} \cite{li_out_2018} and \citeauthor{howe2017engineering} \cite{howe2017engineering}) or web platforms (following \citeauthor{zhang2014wedo} \cite{zhang2014wedo}) that help them participate in coordinated data poisoning with sophisticated means of producing poisoned data (e.g. \cite{shan2020fawkes,Fang2020Apr}). For instance, one could imagine a data poisoning platform, modeled on existing social computing platforms \cite{kraut_building_2012}, that provides users with bespoke poisoned data that they can contribute to a data poisoning attack. In such a platform, users could upload images poisoned with pixel-level manipulation to spoof image recognition systems, or take suggestions of content to interact with so as to fool recommender systems.

``Data poisoners'' might even take inspiration from recent research on what are known as ``adversarial evasion attacks'' \cite{shan2020fawkes}, attacks that help users protect their own images from facial recognition systems (i.e. ``evade'' the system \cite{pitropakis_taxonomy_2019}). \citeauthor{shan2020fawkes} show that their tool, Fawkes, can imperceptibly alter images so that state-of-the-art facial recognition cannot recognize the altered images \cite{shan2020fawkes}. Such tools might be adapted for data poisoning purposes. 


\subsubsection{How can Data Poisoning be Effective?}
There are several reasons to believe data poisoning might be a powerful source of data leverage. Recent work on sophisticated data poisoning suggests that very small amounts of poisoned data (e.g. using less than 1\% of a training set in work from \citeauthor{geiping2020witches} \cite{geiping2020witches}, using 3\% of a training set in work from \citeauthor{steinhardt2017certified} \cite{steinhardt2017certified}) can meaningfully change the performance of a classifier. Even unsophisticated data poisoning (e.g. playing music one does not actually enjoy) by a majority of users could so completely poison a dataset as to make it unusable.

Progress in adversarial ML could actually end up reducing the public's poisoning-based data leverage, in which case non-poisoning data levers would become more important. Fundamentally, data poisoners are engaging in a contest with data scientists. This means any data poisoning technique runs the risk of becoming outdated --- if a company's data scientists find or invent a defense, the public might lose leverage \cite{steinhardt2017certified,shan2020fawkes}. 

Another interesting outcome of data poisoning is its potential conversion to a data strike. In the case where an organization can detect and delete poisoned data, data poisoning reduces to a data strike. Detectable data poisoning could even be used to replicate a deletion-based data strike. For instance, search engine users could use data poisoning tools such as AdNauseum \cite{howe2017engineering} --- which clicks all ads in a user's browser --- to effectively make their ad click data useless, forcing the search engine operator to delete it.

In general, to harm a tech company, data poisoning involves deception and requires affecting the experiences of other users of a platform.
Consider someone who lies on a dating site, a surprisingly common phenomenon \cite{hancock2007truth,toma2010reading}. The user may protect their privacy, but will also poison their own recommendations (e.g. for romantic partners) and make others' dating experiences worse off. The same logic applies to recommendations for friends, videos, and other goods.

A critical challenge for data leverage will be navigating ethical and legal challenges around when data poisoning is acceptable ~\cite{10.1145/2858036.2858060,brunton_obfuscation_2015,shan2020fawkes,geiping2020witches}. Whether a particular instance of poisoning is interpreted to be political dissidence or sabotage depends on the society where it is enacted and on case-by-case specifics. For instance, in some cases existing laws around computer abuse or fraud may come into play, such as the United States' Computer Fraud and Abuse Act (CFAA)~\cite{cfaa,hunt2015gaming}. 


\subsection{Conscious Data Contribution}
The above tactics operate by harming, or threatening to a harm, a given data-dependent technology. However, there are cases for which harmful tactics are not a good fit. For instance, perhaps users do not have the regulatory support needed to delete past data \cite{Waddell2020Oct} or a new technique for detecting poisoned data foils their poisoning attack. Harmful tactics may also be undesirable because an organization's technologies may actively provide benefits to others (e.g. a ML model that is well known to improve accessibility outcomes).

``Conscious data contribution'' (CDC) \cite{vincent_can_2020} is a promising alternative to harm-based data leverage. In CDC, instead of deleting, withholding, or poisoning data, people give their data to an organization that they support to increase market competition as a source of leverage. People using CDC for data leverage are similar to people engaging in ``political consumption'' \cite{koos_what_2012}, but instead of voting with their wallet, they vote with their data. An exciting aspect of CDC is that while small data strikes struggle to put a dent in large-data technologies because of diminishing returns, CDC by a small group of users takes advantage of diminishing returns and provide a competitor with a large boost in performance. We return to this point later in our assessment of data levers.

\subsubsection{CDC Variants}
Variants of CDC closely mirror variants of data strikes because CDC in a sense is the inverse of data strikes --- where data strikes take, CDC gives. 

The easiest way to engage in CDC is to simply start using another technology with the intention of producing useful data for the organization that operates the technology. Sometimes, these CDC campaigns may also involve a data strike if a user moves from one platform to another, for example abandoning Google and moving to DuckDuckGo. 

In jurisdictions where data portability laws \cite{PortGDPR} require that companies allow users to download their data, users can engage in CDC by downloading data from a target organization and contributing it to the organization's competitor. Many services already allow users to download or otherwise access some of their data contributions, but the usefulness of currently exportable data to other companies remains to be seen \cite{James2020Jan}. 

Similarly to how coordinated data strikes and data poisoning might seek to hurt a particular aspect of a technology, coordinated CDC can enhance specific aspects of a technology's capabilities. In a coordinated CDC campaign, organizers might instruct participants to donate specific types of data, or organizers might seek out specific people to join a campaign, in an effort to focus on contributions towards a specific goal. For instance, in the recommendation context, CDC leaders might seek out comedy movies fans to contribute data to a comedy movie recommender, instead of trying to solicit data about every movie genre. Recommender system researchers have shown that allowing users to filter out their old data could actually improve recommendations \cite{wen2018exploring}, so CDC participants could even use filtering to further target their data contributions.

The idea of CDC has complex relationships with various proposals for ``data markets'' \cite{acemoglu2019too,jia2019towards}, which are designed to give people the ability to sell data that they generate. While data markets allow users to participate in a form of CDC by giving them choices about to whom they will sell data, people may prioritize their personal economic incentives over attempts to gain leverage. A major issue with CDC via data markets is the fact that any data with a social component often has information about more than one person\cite{ben-shahar_data_2019,acemoglu2019too}, which could make it legally and ethically tricky to handle data via markets.


\subsubsection{What Does CDC Look Like in Practice?}
As mentioned above, providing data to online platforms can be a form of Conscious Data Contribution if users aim to increase the performance of these technologies relative to their competitors. As such, there are many existing examples of what CDC might look like in practice. 

Cases in which users switch platforms provide one set of examples. In 2015, many Reddit users expressed dissatisfaction with the platform and eventually migrated to alternative platforms such as Voat and Snapzu \cite{newell2016user}. In doing so, these users performed an act of CDC, explicitly supporting Reddit's competitors. Past work suggests that migrations are an especially likely form of CDC, because an individual user's choice to move platforms as part of a CDC campaign  may lead to people in the user's social network also migrating~\cite{garcia2013social, kumar2011understanding}. Where social networks create friction against data strikes, they can help to drive CDC.

Many research initiatives involve collecting volunteered data, which in certain cases could provide opportunities for CDC. In \citeauthor{silva2020facebook}'s study, people contributed data about their Facebook political ads to researchers for monitoring and auditing purposes~\cite{silva2020facebook}. While research studies on their own are not necessarily CDC (though they could be, if the research helps support competitive data-driven technologies), they can often provide a good example of how CDC might be implemented.

Other types of data sharing and generation can also be CDC. For instance, the ``data donation'' concept explored in the context of data ethics \cite{prainsack_data_2019} could be used for CDC. In some cases participation in human computation~\cite{quinn_human_2011}, crowdsourcing systems, and other social computing platforms ~\cite{kraut_building_2012} could qualify as CDC. For example, under our definition, people who choose to contribute data to protein-folding games could be engaging in a form of CDC \cite{eiben2012increased}, with the potential to exert leverage against other organizations that benefit from protein folding models.



\subsubsection{How Can CDC Be Most Effective?}
CDC has a lower barrier to entry than data strikes and data poisoning because it is possible to engage in CDC without completely stopping use of an existing technology. Despite this advantage, a critical question for any CDC effort will be how much leverage ``helping a competitor'' exerts on the target. For instance, a group of CDC users might be able to successfully improve the ML technologies of a small startup that is competing with a major platform. However, even with improved data-driven technologies, other factors like access to capital and switching costs for users might prevent the startup from competing effectively with the original target of leverage, thus reducing the chance that the original target changes their practices.
In some cases, standing up a viable competitor that has better practices could be the end goal of a CDC campaign, even if does not directly harm another company. By supporting a new viable contender, CDC participants can effectively change the overall relationship between the public and technology companies.

Like data strikes, a key determinant of the effectiveness of CDC will be the level of participation. The more people that participate in CDC, the more powerful it will become, and the degree of effectiveness can be estimated using ML findings and methods as we discuss below. A critical distinction between data strikes and CDC is that while small data strikes may struggle to escape the flat portion of ML learning curves, CDC by a small group can actually provide a huge boost in ML performance to a small organization. We expand on this comparison in the following Assessment section.


%% file: 5.assess.tex
\section{Assessing Data Levers}
In this section, we use three axes to evaluate strengths and weaknesses of each data lever: the \textit{barrier-to-entry} to use a data lever, how \textit{ethical and legal considerations} might complicate the use of a data lever, and finally the \textit{potential impact} of each data lever. Table 2 contains a brief summary of our assessments.


\begin{table*}[]
\caption{Summary of key points from our assessment of data levers.}
\begin{tabular}{llll}
\hline
\textbf{Data Lever}                                                   & \textbf{Barriers to Entry}                                                                                                              & \textbf{Legal and Ethical Considerations}                                                                                                                   & \textbf{Potential Impact}                                                                                                                           \\ \hline
Data Strike                                                           & \begin{tabular}[c]{@{}l@{}}\textit{moderate}:\\ -non-use is challenging\\ -hurts participating users\\ -need for privacy tools\end{tabular}       & \begin{tabular}[c]{@{}l@{}} \textit{lower}:\\ -need privacy laws to delete data\\ -harming tech may be undesirable \\ ~ \end{tabular}                                 & \begin{tabular}[c]{@{}l@{}}\textit{moderate}:\\ -small group has small effect\\ -large group can have huge impact \\ ~ \end{tabular}                                 \\ \hline
Data Poisoning                                                        & \begin{tabular}[c]{@{}l@{}}\textit{higher}:\\ -time/effort/bandwidth costs\\ -may require ML knowledge\\ -may require extra coordination\end{tabular} & \begin{tabular}[c]{@{}l@{}}\textit{higher}:\\ -potentially illegal\\-harming tech may be undesirable\\ -inherently deceptive\end{tabular} & \begin{tabular}[c]{@{}l@{}}\textit{moderate}:\\ -small group can have huge effects\\ -if caught, "reduces" to a strike\\ -constant arms race\end{tabular} \\ \hline
\begin{tabular}[c]{@{}l@{}}Conscious Data\\ Contribution\end{tabular} & \begin{tabular}[c]{@{}l@{}}\textit{lower}:\\ -can continue using existing\\   tech \\ ~ \end{tabular}                                                    & \begin{tabular}[c]{@{}l@{}}\textit{moderate}:\\ -potential to improve harmful technologies\\ -privacy concerns of sharing data \\ ~ \end{tabular}                             & \begin{tabular}[c]{@{}l@{}}\textit{moderate}:\\ -small group can have large effects\\ -large group faces diminishing\\ returns\end{tabular}                   
\end{tabular}
\end{table*}

\subsection{Barriers to Entry}
In general, CDC has the lowest barrier to entry of the data levers we identified. This is because CDC does not require stopping or changing the use of existing technologies, which prior work discussed above indicates can be challenging
(e.g. \cite{baumer2013limiting,baumer2015missing,schoenebeck_giving_2014}). 
A person can continue using existing technologies operated by an organization against which they want to exert leverage while engaging in CDC~\cite{jones_nonrivalry_2019,vincent_can_2020}. The main barriers to transfer-based CDC are regulatory and technical. Do laws help people transfer their data \cite{PortGDPR} and do tools exist to make data transfer realistic?

The barriers to entry for data strikes are more substantial then those for CDC and less substantial than those for data poisoning. While participating in a data strike disrupts a user's access to online platforms, strikes do not necessarily force a user to stop using a platform like a traditional boycott would. For instance, a user who relies on Facebook to communicate with family members could stop engaging with sponsored content on Facebook but continue messaging their family members. An Amazon user might continue buying products but stop leaving ratings and reviews. 
An important downside of data strikes is that they hurt the performance of technologies for participating users. By cutting off data contributions, an individual often reduces their own ability to benefit from a system. As discussed by \citeauthor{vincent_data_2019} \cite{vincent_data_2019}, the effect of a data strike will almost always be most pronounced on the strike participants. 

The barriers to entry for each data lever are also contingent on the bandwidth available to potential participants and any potential data caps or data charges they have.  Data strikes are likely the least limited by bandwidth (although striking against an Internet provider, e.g. Facebook Free Basics, could be challenging~\cite{sen2017inside}). In places where the Internet is easy to access and has relatively high data caps, poisoning data by letting music stream for hours or actively manipulating multimedia may be accessible. In contrast, in places where Internet access is limited~\cite{dye2017locating}, poisoning data may be difficult if not impossible. Similar dynamics likely will apply to CDC: data caps could stifle efforts to engage in CDC.

Many of the barriers to entry discussed above are not equally distributed across different populations, and this means that different populations likely have differing access to data leverage. For instance, with regards to data poisoning, the time available to expend the necessary effort and/or the skills necessary to do so will limit the ability of many  populations to engage in data poisoning. Those most positioned to perform data poisoning attacks are ML researchers, technologists, and others with strong technical skills, an already relatively privileged group. Nonetheless, members of this group could use their powerful position for the benefit of people without these advantages (there is precedent of tech worker organizing along these lines~\cite{TechWorkers}).

Turning to coordination, data leverage campaigns will differ in their coordination needs, with greater coordination requirements raising the barrier to entry for all three data levers. Large-scale data leverage is possible without formal organization: boycotts using Twitter hashtags provide real-world examples \cite{jackson2020hashtagactivism}. However, certain data levers require especially well-coordinated effort to see impact, e.g. sophisticated data poisoning \cite{Fang2020Apr}.

\subsection{Legal and Ethical Considerations}
Data leverage organizers may face legal and ethical challenges. Withholding-based data strikes face the fewest of these challenges. These data strikes require almost no regulatory support as users can simply cease using platforms (keeping in mind the differential barrier to entry concerns discussed above). Deletion-based data strikes require a right to deletion and a guarantee that companies are not \textit{data laundering} by retaining model weights trained on old data \cite{lyons_ftc_2021}. 


The legality of data poisoning is likely to remain an open question, and interdisciplinary work between computer scientists and legal scholars will be critical to understand the legal viability of data poisoning as a type of data leverage (and to do so in different jurisdictions). Arguments about the ethics of obfuscation (which itself can be a form of data poisoning) raised by \citeauthor{brunton_obfuscation_2015} apply directly to the use of all types of data poisoning \cite{brunton_obfuscation_2015}. Participants must contend with the potential effects of dishonesty, wastefulness, and other downstream effects of data poisoning. For instance, there are many harms that could stem from poisoning systems that improve accessibility, block hate speech, or support medical decision-making. 

Interesting legal and ethical questions also emerge around CDC. Notably, if a certain data-driven technology is fundamentally harmful and no version of it can meaningful reduce harms (as can be argued for e.g. certain uses of facial recognition \cite{abebe_roles_2020,gebru2019oxford}), CDC will effectively be neutralized.


Another challenge specific to CDC is that there is the potential that data contributions by one person might violate the privacy of others, as data is rarely truly ``individual''~\cite{acemoglu2019too,ben-shahar_data_2019}. For instance, genetic data about one individual may reveal attributes about their family, while financial data may reveal attributes about their friends. On the legal front, CDC often requires either regulatory support in the form of data portability laws or data export features from tech companies.
\subsection{Potential Impact}
Data strikes and data poisoning harm data-dependent technologies, while CDC improves the performance of a data-dependent technology that can then compete with the technology that is the target of data leverage. We can measure potential impact in terms of performance improvement/degradation, as well as downstream effects (e.g. performance degradation leads to users leaving a platform). Ultimately, we are interested in how likely a data lever is to successfully change an organization's behavior with regards to the goals of the data leverage effort, e.g. making changes related to economic inequality, privacy, environmental impact, technologies that reinforce bias, etc.

A relevant finding from prior work \cite{vincent_mapping_2019} describes how data strikes interact with diminishing returns of data. ML performance exhibits diminishing returns; in general, for a particular task, a system can only get so accurate even with massive increases in available data. As such, when an organization accumulates a sufficient amount of data and begins to receive diminishing returns from new data, that organization is not very vulnerable to small data strikes. Such strikes will --- broadly speaking --- only unwind these diminishing marginal returns. To a company with billions of users, a (relatively) small data strike simply may not matter.

The potential impact of data poisoning is also enormous: a large-scale data poisoning attack could render a dataset completely unusable. This approach is also appealing for bargaining: a group could poison some data contributions, and make some demand in return for the ``antidote''. However, the enormous corporate interest in detecting data poisoning means that the would-be poisoners face a constant arms race with operators of targeted technologies. In the worst case scenario, they will be caught, their poisoned data deleted, and the end effect will be equivalent to a data strike.

CDC campaigns, which improve technology performance, operate in the opposite direction of data strikes. Small-scale CDC could be high impact: about 20\% of the users of a system could help a competitor get around 80\% of the best-case performance \cite{vincent_can_2020}. 
On the other hand, once returns begin to diminish, the marginal effect of additional people engaging in CDC begins to fall. 


Given the current evidence, we believe that the data levers we described have a place in the tool belt of those seeking to change the relationship between tech companies and the public. A critical challenge for data leverage researchers will be identifying the correct tool for a specific job. Based on the technologies a target organization uses, a realistic estimate of how many people might participate in data leverage, and knowledge about the resources available to participants, which data lever is most effective?



%% file: 6.discuss.tex
\section{Discussion}
In this section, we discuss questions associated with data leverage that lie beyond the bounds of our current framework. We first discuss the key question of who might expect to benefit from data leverage, and highlight how data leverage might backfire. Next, we summarize key opportunities for researchers, particularly those working in or around FAccT topics. Finally, we summarize opportunities for policy that can amplify and augment data leverage.

\subsection{Who Benefits from Data Leverage?}
Researchers, practitioners, activists, policymakers and others interested in studying, supporting, or amplifying data leverage to reduce power imbalances must contend with unequal access to data leverage. As discussed above, there is strong reason to expect that inequalities in access to data leverage mirror known patterns in access to technology and other sources of power more generally~\cite{abebe_roles_2020}. However, our framework suggests that data poisoning and CDC in particular might allow small groups to have disproportionate impacts. A group of users with needs not currently met by existing technologies might engage in CDC to support a competitor to existing tech companies, or use sophisticated data poisoning techniques that require coordination and knowledge, but not mass participation. Researchers can play an active and critical role by developing tools and promoting policy that widely distributes the ability to participate in data leverage efforts and receive benefits from data leverage. Future work may also need to contend with the possibility of organizations counteracting data leverage, e.g. removing access to publicly available data to maintain a dominant market position.

\subsection{Data Leverage and Data in the Commons}
Many lucrative data-dependent technologies rely on ``commons'' data (e.g. Wikipedia and OpenStreetMap) in addition to the largely proprietary types of data we have discussed so far (e.g. interaction data, rating data). The same is largely true for a variety of data sources that are privately-owned but are a sort of de facto commons for many purposes (e.g. Reddit data, public Twitter posts). Examples of commons-dependent technologies include large language models (e.g. \cite{brown2020language}), search engines (e.g. \cite{vincent2019measuring,mcmahon_substantial_2017,vincent2020deeper}), and a variety of geographic technologies (e.g. \cite{10.1145/2858036.2858123}). Commons datasets have also been instrumental to the advancement of ML research (e.g.\cite{baumgartner2020pushshift,BibEntry2020Sep}).

How can we view the widespread dependence on commons datasets through the lens of data leverage? Adopting a narrow perspective, all three data levers can certainly be employed using data in the commons. In fact, doing so might be a very effective way of exerting data leverage against a large number of data-dependent technologies at once. For instance, through poisoning (i.e. vandalizing) Wikipedia, one can negatively affect a wide variety of Wikipedia-dependent technologies including Google Search, Bing, and Siri \cite{vincent2019measuring,mcmahon_substantial_2017,vincent2020deeper}. Indeed, this has already been done with humorous intent a number of times (e.g. \cite{wilmhoff_tom_2017}). One could similarly imagine organizing a ``data strike'' of sorts in Wikipedia or OpenStreetMap that sought to ensure that a certain type of information does not appear in these datasets. 

That said, from a broader perspective, it is very likely that data poisoning and data strikes using commons data will do substantially more harm than good. For instance, a concerted effort to vandalize (i.e. poison) Wikipedia will cause substantial damage: it would harm Wikipedia readers across the world and would affect technologies operated by non-targeted organizations in addition to those operated by targeted ones. A similar case could be made for most data strikes.

CDC in the context of commons data presents a more complex set of considerations. Indeed, contributing to a commons dataset like Wikipedia can in some ways be understood as a type of CDC as it helps smaller organizations as well as larger ones. However, an important consideration here is that the ability to make use of commons datasets in data-driven technologies is gated by capital. A salient example is GPT-3, OpenAI's high-profile language model that uses training data from sources like Wikipedia and Reddit \cite{brown2020language}. The unprecedented computing power needed to train GPT-3 highlights how the data labor that improves Wikipedia and Reddit can disproportionately benefit organizations with enormous resources. An unfortunate reinforcing dynamic regarding data leverage and commons data thus emerges: while a huge number of organizations and individuals stand to be harmed by any sort of poisoning attack or strike on commons data, large and wealthy firms often stand to benefit disproportionately from improvements to these data. Future work that focuses on efficient training, smaller models, and related goals can help to mitigate this particular concern. Similarly, efforts to open-source models themselves (e.g. share model weights) could also help.


\subsection{Can Data Leverage Research Backfire?}
We have presented data leverage as a means to empower the public to address concerns around computing systems that exacerbate power imbalances and create negative societal outcomes. 
However, research, tools, and policy intended to help data leverage achieve these goals could do the opposite by empowering groups to perpetuate inequalities and, therefore, achieve socially harmful outcomes. For instance, hate groups take advantage of ``data voids'' in search engines to engage in what can be understood as data poisoning attacks by inserting hateful content and influencing model development \cite{golebiewski2019data}. Why wouldn't these groups also try to use other types of data leverage for similar ends?

There are no clear-cut ways to eliminate these risks, but there are steps that data leverage researchers can take to avoid a ``backfire'' outcome. When designing tools to support data leverage, designers might consider heuristic preventative design from \citeauthor{li_out_2018} \cite{li_out_2018} and try to make harmful uses of a technology more challenging. For instance, a data poisoning tool might only help users poison certain types of images known to be important to a particular company or technology. Designers should also consider the principles of data feminism \cite{cifor2019feminist,garcia2020no}, including those that emphasize challenging existing hierarchies, embracing pluralism and context, and making labor visible. 

\subsection{Key Research Opportunities for Data Leverage}
The concept of data leverage presents exciting research opportunities for many fields. Researchers in FAccT, ML, HCI, STS and related areas in particular have unique opportunities to amplify data leverage. 

Data leverage presents a new way of exerting pressure on corporations to make important changes. Most relevant to the FAccT community, this might involve exerting leverage so that a tech company stops the use of a harmful algorithm \cite{kulynych_pots_2020}, or pushing for new economic relationships between data contributors and AI operators in which the benefits of AI are shared more broadly \cite{posner_radical_2018,vincent_mapping_2019}. Data leverage thus presents a novel avenue for researchers to actively pursue pro-social research roles and goals~\cite{abebe_roles_2020}.

There is enormous potential to support data leverage with ML research methods. Using simulations and small-scale experiments, future work could build a catalog of results that activists could draw on to make predictions about the effectiveness of a particular data lever in a particular context, such as ``if we get $x$ participants to engage in a data strike against technology $y$, we can expect to bring down the accuracy of technology by $z$\%, which will likely be enough to encourage company $c$ to make the changes we are demanding''. As data leverage becomes more mainstream, there may also be opportunities to study real-world examples and answer key questions such as: What are the downstream effects on revenue, user retention, and actual changes in company behavior?

Future design work could build upon the collective action literature and develop tools to coordinate efforts to use data leverage. For example, because collective action’s progress is often opaque to individual participants and this can negatively impact engagement, future work may adopt tactics from ``boycott-assisting technologies'' \cite{li_out_2018} and display the impact of the public’s data strike or poisoning (e.g. this technology has lost 3\% of data). Such tools could also support automating data strikes or data poisoning, similar to AdNauseam \citeauthor{howe2017engineering}, to lower the barrier to entry for the public.  

In addition to data strikes and poisoning, the computing community can also support CDC by addressing data compatibility and portability issues across platforms and technologies. Data generated by users are often highly platform- and/or technology-dependent. For example, ratings for the same restaurant or hotel may vary significantly across review platforms \cite{eslami2017careful, li_rating_2020}. Directly transferring data from one technology to another as an act of CDC may run into compatibility issues and even negatively affect the recipient's performance. There is a need for researchers and practitioners to develop software that automatically translates data generated using one technology into data can truly benefit another technology to maximize the success of CDC-based approaches.

Researchers should also seek to better understand the full set of societal impacts that would result from the widespread use of data leverage. As we have discussed above, we hypothesize that the direct effects of actioning data leverage will often involve broadly positive societal impacts, e.g. improved privacy, better distribution of the economic benefits from AI systems, more democratic governance of AI systems. However, the second- and greater-order effects of these changes are more difficult to assess, and even some direct effects may be negative in some cases as highlighted previously. More generally, data leverage defines a pathway to altering power structures in the current computing paradigm. Alterations of power structures in such a complex sociotechnical environment will almost certainly lead to complex outcomes, and more research will be needed to understand these potential outcomes.

\subsection{Key Policy Opportunities for Data Leverage}
Data leverage stands to benefit heavily from regulatory support. As such, data leverage research should be deeply engaged with policy by highlighting regulatory approaches that are likely to amplify the power of data leverage and address its potential negative impacts. Our taxonomy only scratches the surface of how policy may support data leverage; we are excited for this important direction of future work.

Following directly from our assessment of data levers above, we suggest a variety of ways policy can support data leverage:

\begin{itemize}
    \item Data portability laws will directly enhance CDC, enabling users to contribute data they helped generate in the past.
    \item Right-to-delete laws will enhance data strikes, assuming these laws also account for the possibility that companies might ``launder'' deleted data in model weights.
    \item Data transparency laws that make data collection more apparent may help foster support for data leverage movements.
\end{itemize}

We note that these policy suggestions are generally aligned with policy aimed at addressing privacy concerns. This suggests a potential ``win-win'' situation, in which policy simultaneously supports consumer privacy and enhances data leverage.

Expanding on the above points about data portability and right-to-delete laws, policy also offers the potential for making it easy for individuals to use multiple data levers in conjunction with one another. As mentioned above, there are natural connections between data strikes and CDC: by moving from one platform to a new platform, a user can take advantage of both data levers. However, through regulatory support, it may be possible to engage in much more elaborate combinations of data strikes and CDC, for instance deleting only certain pieces of data and transferring over other pieces of data.

\section{Conclusion}
In this paper, we presented a framework for using ``data leverage'' to give the public more influence over technology company behavior. Drawing on a variety of research areas, we described and assessed the ``data levers'' available to the public. We highlighted key areas where researchers and policymakers can amplify data leverage and work to ensure data leverage distributes power more broadly than is the case in the status quo.

\section{Acknowledgments}
This work was funded in part by NSF grants 1815507 and 1707296. We are grateful for feedback from colleagues at the CollabLab at Northwestern, GroupLens at the University of Minnesota, and the Community Data Science Collective.

%% file: 0.paper.bbl

\begin{thebibliography}{136}


\ifx \showCODEN    \undefined \def \showCODEN     #1{\unskip}     \fi
\ifx \showDOI      \undefined \def \showDOI       #1{#1}\fi
\ifx \showISBNx    \undefined \def \showISBNx     #1{\unskip}     \fi
\ifx \showISBNxiii \undefined \def \showISBNxiii  #1{\unskip}     \fi
\ifx \showISSN     \undefined \def \showISSN      #1{\unskip}     \fi
\ifx \showLCCN     \undefined \def \showLCCN      #1{\unskip}     \fi
\ifx \shownote     \undefined \def \shownote      #1{#1}          \fi
\ifx \showarticletitle \undefined \def \showarticletitle #1{#1}   \fi
\ifx \showURL      \undefined \def \showURL       {\relax}        \fi
\providecommand\bibfield[2]{#2}
\providecommand\bibinfo[2]{#2}
\providecommand\natexlab[1]{#1}
\providecommand\showeprint[2][]{arXiv:#2}

\bibitem[\protect\citeauthoryear{??}{Por}{2018}]%
        {PortGDPR}
 \bibinfo{year}{2018}\natexlab{}.
\newblock \bibinfo{title}{{Art. 20 GDPR {\textendash} Right to data portability
  {$\vert$} General Data Protection Regulation (GDPR)}}.
\newblock
\newblock
\urldef\tempurl%
\url{https://gdpr-info.eu/art-20-gdpr}
\showURL{%
\tempurl}


\bibitem[\protect\citeauthoryear{??}{cfa}{2020}]%
        {cfaa}
 \bibinfo{year}{2020}\natexlab{}.
\newblock \bibinfo{title}{{18 U.S. Code {\ifmmode\S\else\textsection\fi} 1030 -
  Fraud and related activity in connection with computers}}.
\newblock
\newblock
\urldef\tempurl%
\url{https://www.law.cornell.edu/uscode/text/18/1030}
\showURL{%
\tempurl}
\newblock
\shownote{[Online; accessed 7. Oct. 2020].}


\bibitem[\protect\citeauthoryear{??}{Tec}{2020}]%
        {TechWorkers}
 \bibinfo{year}{2020}\natexlab{}.
\newblock \bibinfo{title}{{Tech Workers Coalition}}.
\newblock
\newblock
\urldef\tempurl%
\url{https://techworkerscoalition.org}
\showURL{%
\tempurl}
\newblock
\shownote{[Online; accessed 6. Oct. 2020].}


\bibitem[\protect\citeauthoryear{??}{Bib}{2020}]%
        {BibEntry2020Sep}
 \bibinfo{year}{2020}\natexlab{}.
\newblock \bibinfo{title}{{Wikipedia:Academic studies of Wikipedia -
  Wikipedia}}.
\newblock
\newblock
\urldef\tempurl%
\url{https://en.wikipedia.org/w/index.php?title=Wikipedia:Academic_studies_of_Wikipedia&oldid=971074694}
\showURL{%
\tempurl}
\newblock
\shownote{[Online; accessed 29. Sep. 2020].}


\bibitem[\protect\citeauthoryear{Abebe, Barocas, Kleinberg, Levy, Raghavan, and
  Robinson}{Abebe et~al\mbox{.}}{2020}]%
        {abebe_roles_2020}
\bibfield{author}{\bibinfo{person}{Rediet Abebe}, \bibinfo{person}{Solon
  Barocas}, \bibinfo{person}{Jon Kleinberg}, \bibinfo{person}{Karen Levy},
  \bibinfo{person}{Manish Raghavan}, {and} \bibinfo{person}{David~G Robinson}.}
  \bibinfo{year}{2020}\natexlab{}.
\newblock \showarticletitle{Roles for computing in social change}. In
  \bibinfo{booktitle}{\emph{Proceedings of the 2020 {Conference} on {Fairness},
  {Accountability}, and {Transparency}}}. \bibinfo{pages}{252--260}.
\newblock


\bibitem[\protect\citeauthoryear{Acemoglu, Makhdoumi, Malekian, and
  Ozdaglar}{Acemoglu et~al\mbox{.}}{2019}]%
        {acemoglu2019too}
\bibfield{author}{\bibinfo{person}{Daron Acemoglu}, \bibinfo{person}{Ali
  Makhdoumi}, \bibinfo{person}{Azarakhsh Malekian}, {and}
  \bibinfo{person}{Asuman Ozdaglar}.} \bibinfo{year}{2019}\natexlab{}.
\newblock \bibinfo{booktitle}{\emph{Too much data: Prices and inefficiencies in
  data markets}}.
\newblock \bibinfo{type}{{T}echnical {R}eport}. \bibinfo{institution}{National
  Bureau of Economic Research}.
\newblock


\bibitem[\protect\citeauthoryear{Albert, Penney, Schneier, and
  Siva~Kumar}{Albert et~al\mbox{.}}{2020}]%
        {albert2020politics}
\bibfield{author}{\bibinfo{person}{Kendra Albert}, \bibinfo{person}{Jon
  Penney}, \bibinfo{person}{Bruce Schneier}, {and} \bibinfo{person}{Ram~Shankar
  Siva~Kumar}.} \bibinfo{year}{2020}\natexlab{}.
\newblock \showarticletitle{Politics of Adversarial Machine Learning}. In
  \bibinfo{booktitle}{\emph{Towards Trustworthy ML: Rethinking Security and
  Privacy for ML Workshop, Eighth International Conference on Learning
  Representations (ICLR)}}.
\newblock


\bibitem[\protect\citeauthoryear{Arrieta~Ibarra, Goff, Jiménez~Hernández,
  Lanier, and Weyl}{Arrieta~Ibarra et~al\mbox{.}}{2018}]%
        {arrieta_ibarra_should_2018}
\bibfield{author}{\bibinfo{person}{Imanol Arrieta~Ibarra},
  \bibinfo{person}{Leonard Goff}, \bibinfo{person}{Diego Jiménez~Hernández},
  \bibinfo{person}{Jaron Lanier}, {and} \bibinfo{person}{E Weyl}.}
  \bibinfo{year}{2018}\natexlab{}.
\newblock \showarticletitle{Should {We} {Treat} {Data} as {Labor}? {Moving}
  {Beyond} '{Free}'}.
\newblock \bibinfo{journal}{\emph{American Economic Association Papers \&
  Proceedings}} \bibinfo{volume}{1}, \bibinfo{number}{1}
  (\bibinfo{year}{2018}).
\newblock


\bibitem[\protect\citeauthoryear{Baack}{Baack}{2015}]%
        {Baack}
\bibfield{author}{\bibinfo{person}{Stefan Baack}.}
  \bibinfo{year}{2015}\natexlab{}.
\newblock \showarticletitle{Datafication and empowerment: How the open data
  movement re-articulates notions of democracy, participation, and journalism}.
\newblock \bibinfo{journal}{\emph{Big Data \& Society}} \bibinfo{volume}{2},
  \bibinfo{number}{2} (\bibinfo{year}{2015}),
  \bibinfo{pages}{2053951715594634}.
\newblock
\urldef\tempurl%
\url{https://doi.org/10.1177/2053951715594634}
\showDOI{\tempurl}
\showeprint{https://doi.org/10.1177/2053951715594634}


\bibitem[\protect\citeauthoryear{Barreno, Nelson, Sears, Joseph, and
  Tygar}{Barreno et~al\mbox{.}}{2006}]%
        {barreno2006can}
\bibfield{author}{\bibinfo{person}{Marco Barreno}, \bibinfo{person}{Blaine
  Nelson}, \bibinfo{person}{Russell Sears}, \bibinfo{person}{Anthony~D Joseph},
  {and} \bibinfo{person}{J~Doug Tygar}.} \bibinfo{year}{2006}\natexlab{}.
\newblock \showarticletitle{Can machine learning be secure?}. In
  \bibinfo{booktitle}{\emph{Proceedings of the 2006 ACM Symposium on
  Information, computer and communications security}}. \bibinfo{pages}{16--25}.
\newblock


\bibitem[\protect\citeauthoryear{Baumer}{Baumer}{2014}]%
        {baumer2014refusing}
\bibfield{author}{\bibinfo{person}{Ames Morgan G. Brubaker Jed R. Burrell Jenna
  Dourish~Paul Baumer, Eric~PS}.} \bibinfo{year}{2014}\natexlab{}.
\newblock \showarticletitle{Refusing, Limiting, Departing: Why We Should Study
  Technology Non-Use}. In \bibinfo{booktitle}{\emph{CHI EA '14: CHI '14
  Extended Abstracts on Human Factors in Computing Systems}}.
  \bibinfo{pages}{65--68}.
\newblock


\bibitem[\protect\citeauthoryear{Baumer}{Baumer}{2018}]%
        {baumer2018socioeconomic}
\bibfield{author}{\bibinfo{person}{Eric~PS Baumer}.}
  \bibinfo{year}{2018}\natexlab{}.
\newblock \showarticletitle{Socioeconomic Inequalities in the Non use of
  Facebook}. In \bibinfo{booktitle}{\emph{Proceedings of the 2018 CHI
  Conference on Human Factors in Computing Systems}}. \bibinfo{pages}{1--14}.
\newblock


\bibitem[\protect\citeauthoryear{Baumer, Adams, Khovanskaya, Liao, Smith,
  Schwanda~Sosik, and Williams}{Baumer et~al\mbox{.}}{2013}]%
        {baumer2013limiting}
\bibfield{author}{\bibinfo{person}{Eric~PS Baumer}, \bibinfo{person}{Phil
  Adams}, \bibinfo{person}{Vera~D Khovanskaya}, \bibinfo{person}{Tony~C Liao},
  \bibinfo{person}{Madeline~E Smith}, \bibinfo{person}{Victoria
  Schwanda~Sosik}, {and} \bibinfo{person}{Kaiton Williams}.}
  \bibinfo{year}{2013}\natexlab{}.
\newblock \showarticletitle{Limiting, leaving, and (re) lapsing: an exploration
  of facebook non-use practices and experiences}. In
  \bibinfo{booktitle}{\emph{Proceedings of the SIGCHI conference on human
  factors in computing systems}}. \bibinfo{pages}{3257--3266}.
\newblock


\bibitem[\protect\citeauthoryear{Baumer, Guha, Quan, Mimno, and Gay}{Baumer
  et~al\mbox{.}}{2015}]%
        {baumer2015missing}
\bibfield{author}{\bibinfo{person}{Eric~PS Baumer}, \bibinfo{person}{Shion
  Guha}, \bibinfo{person}{Emily Quan}, \bibinfo{person}{David Mimno}, {and}
  \bibinfo{person}{Geri~K Gay}.} \bibinfo{year}{2015}\natexlab{}.
\newblock \showarticletitle{Missing photos, suffering withdrawal, or finding
  freedom? How experiences of social media non-use influence the likelihood of
  reversion}.
\newblock \bibinfo{journal}{\emph{Social Media+ Society}} \bibinfo{volume}{1},
  \bibinfo{number}{2} (\bibinfo{year}{2015}),
  \bibinfo{pages}{2056305115614851}.
\newblock


\bibitem[\protect\citeauthoryear{Baumer, Guha, Skeba, and Gay}{Baumer
  et~al\mbox{.}}{2019}]%
        {baumer2019all}
\bibfield{author}{\bibinfo{person}{Eric~PS Baumer}, \bibinfo{person}{Shion
  Guha}, \bibinfo{person}{Patrick Skeba}, {and} \bibinfo{person}{Geraldine
  Gay}.} \bibinfo{year}{2019}\natexlab{}.
\newblock \showarticletitle{All Users are (Not) Created Equal: Predictors Vary
  for Different Forms of Facebook Non/use}.
\newblock \bibinfo{journal}{\emph{Proceedings of the ACM on Human-Computer
  Interaction}} \bibinfo{volume}{3}, \bibinfo{number}{CSCW}
  (\bibinfo{year}{2019}), \bibinfo{pages}{1--28}.
\newblock


\bibitem[\protect\citeauthoryear{Baumgartner, Zannettou, Keegan, Squire, and
  Blackburn}{Baumgartner et~al\mbox{.}}{2020}]%
        {baumgartner2020pushshift}
\bibfield{author}{\bibinfo{person}{Jason Baumgartner}, \bibinfo{person}{Savvas
  Zannettou}, \bibinfo{person}{Brian Keegan}, \bibinfo{person}{Megan Squire},
  {and} \bibinfo{person}{Jeremy Blackburn}.} \bibinfo{year}{2020}\natexlab{}.
\newblock \showarticletitle{The pushshift reddit dataset}. In
  \bibinfo{booktitle}{\emph{Proceedings of the International AAAI Conference on
  Web and Social Media}}, Vol.~\bibinfo{volume}{14}. \bibinfo{pages}{830--839}.
\newblock


\bibitem[\protect\citeauthoryear{Ben-Shahar}{Ben-Shahar}{2019}]%
        {ben-shahar_data_2019}
\bibfield{author}{\bibinfo{person}{Omri Ben-Shahar}.}
  \bibinfo{year}{2019}\natexlab{}.
\newblock \showarticletitle{Data {Pollution}}.
\newblock \bibinfo{journal}{\emph{Journal of Legal Analysis}}
  \bibinfo{volume}{11} (\bibinfo{year}{2019}), \bibinfo{pages}{104--159}.
\newblock
\newblock
\shownote{Publisher: Narnia.}


\bibitem[\protect\citeauthoryear{Benjamin}{Benjamin}{2016}]%
        {benjamin_informed_2016}
\bibfield{author}{\bibinfo{person}{Ruha Benjamin}.}
  \bibinfo{year}{2016}\natexlab{}.
\newblock \showarticletitle{Informed refusal: {Toward} a justice-based
  bioethics}.
\newblock \bibinfo{journal}{\emph{Science, Technology, \& Human Values}}
  \bibinfo{volume}{41}, \bibinfo{number}{6} (\bibinfo{year}{2016}),
  \bibinfo{pages}{967--990}.
\newblock
\newblock
\shownote{Publisher: SAGE Publications Sage CA: Los Angeles, CA.}


\bibitem[\protect\citeauthoryear{Biggio, Nelson, and Laskov}{Biggio
  et~al\mbox{.}}{2012}]%
        {biggio2012poisoning}
\bibfield{author}{\bibinfo{person}{Battista Biggio}, \bibinfo{person}{Blaine
  Nelson}, {and} \bibinfo{person}{Pavel Laskov}.}
  \bibinfo{year}{2012}\natexlab{}.
\newblock \showarticletitle{Poisoning attacks against support vector machines}.
\newblock \bibinfo{journal}{\emph{arXiv preprint arXiv:1206.6389}}
  (\bibinfo{year}{2012}).
\newblock


\bibitem[\protect\citeauthoryear{Binns}{Binns}{2018}]%
        {binns2018fairness}
\bibfield{author}{\bibinfo{person}{Reuben Binns}.}
  \bibinfo{year}{2018}\natexlab{}.
\newblock \showarticletitle{Fairness in machine learning: Lessons from
  political philosophy}. In \bibinfo{booktitle}{\emph{Conference on Fairness,
  Accountability and Transparency}}. \bibinfo{pages}{149--159}.
\newblock


\bibitem[\protect\citeauthoryear{Bloodgood and Callison-Burch}{Bloodgood and
  Callison-Burch}{2010}]%
        {bloodgood_bucking_2010}
\bibfield{author}{\bibinfo{person}{Michael Bloodgood} {and}
  \bibinfo{person}{Chris Callison-Burch}.} \bibinfo{year}{2010}\natexlab{}.
\newblock \showarticletitle{Bucking the trend: large-scale cost-focused active
  learning for statistical machine translation}. In
  \bibinfo{booktitle}{\emph{Proceedings of the 48th {Annual} {Meeting} of the
  {Association} for {Computational} {Linguistics}}}.
  \bibinfo{publisher}{Association for Computational Linguistics},
  \bibinfo{pages}{854--864}.
\newblock


\bibitem[\protect\citeauthoryear{Brown, Mann, Ryder, Subbiah, Kaplan, Dhariwal,
  Neelakantan, Shyam, Sastry, Askell, Agarwal, Herbert-Voss, Krueger, Henighan,
  Child, Ramesh, Ziegler, Wu, Winter, Hesse, Chen, Sigler, Litwin, Gray, Chess,
  Clark, Berner, McCandlish, Radford, Sutskever, and Amodei}{Brown
  et~al\mbox{.}}{2020}]%
        {brown2020language}
\bibfield{author}{\bibinfo{person}{Tom~B. Brown}, \bibinfo{person}{Benjamin
  Mann}, \bibinfo{person}{Nick Ryder}, \bibinfo{person}{Melanie Subbiah},
  \bibinfo{person}{Jared Kaplan}, \bibinfo{person}{Prafulla Dhariwal},
  \bibinfo{person}{Arvind Neelakantan}, \bibinfo{person}{Pranav Shyam},
  \bibinfo{person}{Girish Sastry}, \bibinfo{person}{Amanda Askell},
  \bibinfo{person}{Sandhini Agarwal}, \bibinfo{person}{Ariel Herbert-Voss},
  \bibinfo{person}{Gretchen Krueger}, \bibinfo{person}{Tom Henighan},
  \bibinfo{person}{Rewon Child}, \bibinfo{person}{Aditya Ramesh},
  \bibinfo{person}{Daniel~M. Ziegler}, \bibinfo{person}{Jeffrey Wu},
  \bibinfo{person}{Clemens Winter}, \bibinfo{person}{Christopher Hesse},
  \bibinfo{person}{Mark Chen}, \bibinfo{person}{Eric Sigler},
  \bibinfo{person}{Mateusz Litwin}, \bibinfo{person}{Scott Gray},
  \bibinfo{person}{Benjamin Chess}, \bibinfo{person}{Jack Clark},
  \bibinfo{person}{Christopher Berner}, \bibinfo{person}{Sam McCandlish},
  \bibinfo{person}{Alec Radford}, \bibinfo{person}{Ilya Sutskever}, {and}
  \bibinfo{person}{Dario Amodei}.} \bibinfo{year}{2020}\natexlab{}.
\newblock \bibinfo{title}{Language Models are Few-Shot Learners}.
\newblock
\newblock
\showeprint[arxiv]{cs.CL/2005.14165}


\bibitem[\protect\citeauthoryear{Brubaker, Ananny, and Crawford}{Brubaker
  et~al\mbox{.}}{2016}]%
        {brubaker2016departing}
\bibfield{author}{\bibinfo{person}{Jed~R Brubaker}, \bibinfo{person}{Mike
  Ananny}, {and} \bibinfo{person}{Kate Crawford}.}
  \bibinfo{year}{2016}\natexlab{}.
\newblock \showarticletitle{Departing glances: A sociotechnical account of
  ‘leaving’Grindr}.
\newblock \bibinfo{journal}{\emph{New Media \& Society}} \bibinfo{volume}{18},
  \bibinfo{number}{3} (\bibinfo{year}{2016}), \bibinfo{pages}{373--390}.
\newblock


\bibitem[\protect\citeauthoryear{Brunton and Nissenbaum}{Brunton and
  Nissenbaum}{2015}]%
        {brunton_obfuscation_2015}
\bibfield{author}{\bibinfo{person}{Finn Brunton} {and}
  \bibinfo{person}{Helen~Fay Nissenbaum}.} \bibinfo{year}{2015}\natexlab{}.
\newblock \bibinfo{booktitle}{\emph{Obfuscation: a user's guide for privacy and
  protest}}.
\newblock \bibinfo{publisher}{MIT Press}, \bibinfo{address}{Cambridge,
  Massachusetts}.
\newblock
\showISBNx{978-0-262-02973-5}


\bibitem[\protect\citeauthoryear{Brynjolfsson and McAfee}{Brynjolfsson and
  McAfee}{2014}]%
        {brynjolfsson2014second}
\bibfield{author}{\bibinfo{person}{Erik Brynjolfsson} {and}
  \bibinfo{person}{Andrew McAfee}.} \bibinfo{year}{2014}\natexlab{}.
\newblock \bibinfo{booktitle}{\emph{The second machine age: Work, progress, and
  prosperity in a time of brilliant technologies}}.
\newblock \bibinfo{publisher}{WW Norton \& Company}.
\newblock


\bibitem[\protect\citeauthoryear{Brynjolfsson and McElheran}{Brynjolfsson and
  McElheran}{2016}]%
        {brynjolfsson_rapid_2016}
\bibfield{author}{\bibinfo{person}{Erik Brynjolfsson} {and}
  \bibinfo{person}{Kristina McElheran}.} \bibinfo{year}{2016}\natexlab{}.
\newblock \showarticletitle{The rapid adoption of data-driven decision-making}.
\newblock \bibinfo{journal}{\emph{American Economic Review}}
  \bibinfo{volume}{106}, \bibinfo{number}{5} (\bibinfo{year}{2016}),
  \bibinfo{pages}{133--39}.
\newblock


\bibitem[\protect\citeauthoryear{Budak, Goel, Rao, and Zervas}{Budak
  et~al\mbox{.}}{2016}]%
        {budak_understanding_2016}
\bibfield{author}{\bibinfo{person}{Ceren Budak}, \bibinfo{person}{Sharad Goel},
  \bibinfo{person}{Justin Rao}, {and} \bibinfo{person}{Georgios Zervas}.}
  \bibinfo{year}{2016}\natexlab{}.
\newblock \showarticletitle{Understanding {Emerging} {Threats} to {Online}
  {Advertising}}. In \bibinfo{booktitle}{\emph{Proceedings of the 2016 {ACM}
  {Conference} on {Economics} and {Computation}}} \emph{(\bibinfo{series}{{EC}
  '16})}. \bibinfo{publisher}{ACM}, \bibinfo{address}{New York, NY, USA},
  \bibinfo{pages}{561--578}.
\newblock
\showISBNx{978-1-4503-3936-0}
\urldef\tempurl%
\url{https://doi.org/10.1145/2940716.2940787}
\showDOI{\tempurl}
\newblock
\shownote{event-place: Maastricht, The Netherlands.}


\bibitem[\protect\citeauthoryear{Casemajor, Couture, Delfin, Goerzen, and
  Delfanti}{Casemajor et~al\mbox{.}}{2015}]%
        {casemajor_non-participation_2015}
\bibfield{author}{\bibinfo{person}{Nathalie Casemajor},
  \bibinfo{person}{Stıfmmode{\textbackslash}acutee{\textbackslash}elseé{\textbackslash}fiphane
  Couture}, \bibinfo{person}{Mauricio Delfin}, \bibinfo{person}{Matthew
  Goerzen}, {and} \bibinfo{person}{Alessandro Delfanti}.}
  \bibinfo{year}{2015}\natexlab{}.
\newblock \showarticletitle{Non-participation in digital media: toward a
  framework of mediated political action}.
\newblock \bibinfo{journal}{\emph{Media, Culture \& Society}}
  \bibinfo{volume}{37}, \bibinfo{number}{6} (\bibinfo{date}{May}
  \bibinfo{year}{2015}), \bibinfo{pages}{850--866}.
\newblock
\showISSN{0163-4437}
\urldef\tempurl%
\url{https://doi.org/10.1177/0163443715584098}
\showDOI{\tempurl}
\newblock
\shownote{Publisher: SAGE Publications Ltd.}


\bibitem[\protect\citeauthoryear{Chancellor, Baumer, and
  De~Choudhury}{Chancellor et~al\mbox{.}}{2019}]%
        {chancellor2019human}
\bibfield{author}{\bibinfo{person}{Stevie Chancellor}, \bibinfo{person}{Eric~PS
  Baumer}, {and} \bibinfo{person}{Munmun De~Choudhury}.}
  \bibinfo{year}{2019}\natexlab{}.
\newblock \showarticletitle{Who is the" Human" in Human-Centered Machine
  Learning: The Case of Predicting Mental Health from Social Media}.
\newblock \bibinfo{journal}{\emph{Proceedings of the ACM on Human-Computer
  Interaction}} \bibinfo{volume}{3}, \bibinfo{number}{CSCW}
  (\bibinfo{year}{2019}), \bibinfo{pages}{1--32}.
\newblock


\bibitem[\protect\citeauthoryear{Chirita, Nejdl, and Zamfir}{Chirita
  et~al\mbox{.}}{2005}]%
        {chirita_preventing_2005}
\bibfield{author}{\bibinfo{person}{Paul-Alexandru Chirita},
  \bibinfo{person}{Wolfgang Nejdl}, {and} \bibinfo{person}{Cristian Zamfir}.}
  \bibinfo{year}{2005}\natexlab{}.
\newblock \showarticletitle{Preventing shilling attacks in online recommender
  systems}. In \bibinfo{booktitle}{\emph{Proceedings of the 7th annual {ACM}
  international workshop on {Web} information and data management}}.
  \bibinfo{pages}{67--74}.
\newblock


\bibitem[\protect\citeauthoryear{Cho, Lee, Shin, Choy, and Do}{Cho
  et~al\mbox{.}}{2015}]%
        {cho2015much}
\bibfield{author}{\bibinfo{person}{Junghwan Cho}, \bibinfo{person}{Kyewook
  Lee}, \bibinfo{person}{Ellie Shin}, \bibinfo{person}{Garry Choy}, {and}
  \bibinfo{person}{Synho Do}.} \bibinfo{year}{2015}\natexlab{}.
\newblock \showarticletitle{How much data is needed to train a medical image
  deep learning system to achieve necessary high accuracy?}
\newblock \bibinfo{journal}{\emph{arXiv preprint arXiv:1511.06348}}
  (\bibinfo{year}{2015}).
\newblock


\bibitem[\protect\citeauthoryear{Cho}{Cho}{2011}]%
        {cho2011unsell}
\bibfield{author}{\bibinfo{person}{Max Cho}.} \bibinfo{year}{2011}\natexlab{}.
\newblock \showarticletitle{Unsell Yourself—A Protest Model Against
  Facebook}.
\newblock \bibinfo{journal}{\emph{Yale Law \& Technology}}
  (\bibinfo{year}{2011}).
\newblock


\bibitem[\protect\citeauthoryear{Cifor, Garcia, Cowan, Rault, Sutherland, Chan,
  Rode, Hoffmann, Salehi, and Nakamura}{Cifor et~al\mbox{.}}{2019}]%
        {cifor2019feminist}
\bibfield{author}{\bibinfo{person}{Marika Cifor}, \bibinfo{person}{Patricia
  Garcia}, \bibinfo{person}{TL Cowan}, \bibinfo{person}{Jasmine Rault},
  \bibinfo{person}{Tonia Sutherland}, \bibinfo{person}{Anita~Say Chan},
  \bibinfo{person}{Jennifer Rode}, \bibinfo{person}{Anna~Lauren Hoffmann},
  \bibinfo{person}{Niloufar Salehi}, {and} \bibinfo{person}{Lisa Nakamura}.}
  \bibinfo{year}{2019}\natexlab{}.
\newblock \bibinfo{title}{Feminist data manifest-no}.
\newblock
\newblock


\bibitem[\protect\citeauthoryear{Couldry and Powell}{Couldry and
  Powell}{2014}]%
        {couldry2014big}
\bibfield{author}{\bibinfo{person}{Nick Couldry} {and} \bibinfo{person}{Alison
  Powell}.} \bibinfo{year}{2014}\natexlab{}.
\newblock \showarticletitle{Big data from the bottom up}.
\newblock \bibinfo{journal}{\emph{Big Data \& Society}} \bibinfo{volume}{1},
  \bibinfo{number}{2} (\bibinfo{year}{2014}),
  \bibinfo{pages}{2053951714539277}.
\newblock


\bibitem[\protect\citeauthoryear{Crawford and Joler}{Crawford and
  Joler}{2018}]%
        {crawford2018anatomy}
\bibfield{author}{\bibinfo{person}{Kate Crawford} {and} \bibinfo{person}{Vladan
  Joler}.} \bibinfo{year}{2018}\natexlab{}.
\newblock \showarticletitle{Anatomy of an AI System-The Amazon Echo as an
  anatomical map of human labor, data and planetary resources}.
\newblock \bibinfo{journal}{\emph{AI Now Institute and Share Lab}}
  \bibinfo{volume}{7} (\bibinfo{year}{2018}).
\newblock


\bibitem[\protect\citeauthoryear{Dietterich and Kong}{Dietterich and
  Kong}{1995}]%
        {dietterich_machine_1995}
\bibfield{author}{\bibinfo{person}{Thomas~G Dietterich} {and}
  \bibinfo{person}{Eun~Bae Kong}.} \bibinfo{year}{1995}\natexlab{}.
\newblock \bibinfo{booktitle}{\emph{Machine learning bias, statistical bias,
  and statistical variance of decision tree algorithms}}.
\newblock \bibinfo{type}{{T}echnical {R}eport}. \bibinfo{institution}{Technical
  report, Department of Computer Science, Oregon State University}.
\newblock


\bibitem[\protect\citeauthoryear{D'Ignazio and Klein}{D'Ignazio and
  Klein}{2020}]%
        {d2020data}
\bibfield{author}{\bibinfo{person}{Catherine D'Ignazio} {and}
  \bibinfo{person}{Lauren~F Klein}.} \bibinfo{year}{2020}\natexlab{}.
\newblock \bibinfo{booktitle}{\emph{Data feminism}}.
\newblock \bibinfo{publisher}{MIT Press}.
\newblock


\bibitem[\protect\citeauthoryear{Dye, Nemer, Pina, Sambasivan, Bruckman, and
  Kumar}{Dye et~al\mbox{.}}{2017}]%
        {dye2017locating}
\bibfield{author}{\bibinfo{person}{Michaelanne Dye}, \bibinfo{person}{David
  Nemer}, \bibinfo{person}{Laura~R Pina}, \bibinfo{person}{Nithya Sambasivan},
  \bibinfo{person}{Amy~S Bruckman}, {and} \bibinfo{person}{Neha Kumar}.}
  \bibinfo{year}{2017}\natexlab{}.
\newblock \showarticletitle{Locating the Internet in the Parks of Havana}. In
  \bibinfo{booktitle}{\emph{Proceedings of the 2017 CHI Conference on Human
  Factors in Computing Systems}}. \bibinfo{pages}{3867--3878}.
\newblock


\bibitem[\protect\citeauthoryear{Eiben, Siegel, Bale, Cooper, Khatib, Shen,
  Stoddard, Popovic, and Baker}{Eiben et~al\mbox{.}}{2012}]%
        {eiben2012increased}
\bibfield{author}{\bibinfo{person}{Christopher~B Eiben},
  \bibinfo{person}{Justin~B Siegel}, \bibinfo{person}{Jacob~B Bale},
  \bibinfo{person}{Seth Cooper}, \bibinfo{person}{Firas Khatib},
  \bibinfo{person}{Betty~W Shen}, \bibinfo{person}{Barry~L Stoddard},
  \bibinfo{person}{Zoran Popovic}, {and} \bibinfo{person}{David Baker}.}
  \bibinfo{year}{2012}\natexlab{}.
\newblock \showarticletitle{Increased Diels-Alderase activity through backbone
  remodeling guided by Foldit players}.
\newblock \bibinfo{journal}{\emph{Nature biotechnology}} \bibinfo{volume}{30},
  \bibinfo{number}{2} (\bibinfo{year}{2012}), \bibinfo{pages}{190--192}.
\newblock


\bibitem[\protect\citeauthoryear{Ekstrand, Joshaghani, and Mehrpouyan}{Ekstrand
  et~al\mbox{.}}{2018}]%
        {pmlr-v81-ekstrand18a}
\bibfield{author}{\bibinfo{person}{Michael~D. Ekstrand},
  \bibinfo{person}{Rezvan Joshaghani}, {and} \bibinfo{person}{Hoda
  Mehrpouyan}.} \bibinfo{year}{2018}\natexlab{}.
\newblock \showarticletitle{Privacy for All: Ensuring Fair and Equitable
  Privacy Protections} \emph{(\bibinfo{series}{Proceedings of Machine Learning
  Research})}, \bibfield{editor}{\bibinfo{person}{Sorelle~A. Friedler} {and}
  \bibinfo{person}{Christo Wilson}} (Eds.), Vol.~\bibinfo{volume}{81}.
  \bibinfo{publisher}{PMLR}, \bibinfo{address}{New York, NY, USA},
  \bibinfo{pages}{35--47}.
\newblock
\urldef\tempurl%
\url{http://proceedings.mlr.press/v81/ekstrand18a.html}
\showURL{%
\tempurl}


\bibitem[\protect\citeauthoryear{Eskandanian, Sonboli, and
  Mobasher}{Eskandanian et~al\mbox{.}}{2019}]%
        {eskandanian_power_2019}
\bibfield{author}{\bibinfo{person}{Farzad Eskandanian}, \bibinfo{person}{Nasim
  Sonboli}, {and} \bibinfo{person}{Bamshad Mobasher}.}
  \bibinfo{year}{2019}\natexlab{}.
\newblock \showarticletitle{Power of the {Few}: {Analyzing} the {Impact} of
  {Influential} {Users} in {Collaborative} {Recommender} {Systems}}. In
  \bibinfo{booktitle}{\emph{Proceedings of the 27th {ACM} {Conference} on
  {User} {Modeling}, {Adaptation} and {Personalization}}}.
  \bibinfo{pages}{225--233}.
\newblock


\bibitem[\protect\citeauthoryear{Eslami, Vaccaro, Karahalios, and
  Hamilton}{Eslami et~al\mbox{.}}{2017}]%
        {eslami2017careful}
\bibfield{author}{\bibinfo{person}{Motahhare Eslami}, \bibinfo{person}{Kristen
  Vaccaro}, \bibinfo{person}{Karrie Karahalios}, {and} \bibinfo{person}{Kevin
  Hamilton}.} \bibinfo{year}{2017}\natexlab{}.
\newblock \showarticletitle{" Be Careful; Things Can Be Worse than They
  Appear": Understanding Biased Algorithms and Users' Behavior Around Them in
  Rating Platforms.}. In \bibinfo{booktitle}{\emph{ICWSM}}.
  \bibinfo{pages}{62--71}.
\newblock


\bibitem[\protect\citeauthoryear{Eslami, Vaccaro, Lee, Elazari Bar~On, Gilbert,
  and Karahalios}{Eslami et~al\mbox{.}}{2019}]%
        {eslami2019user}
\bibfield{author}{\bibinfo{person}{Motahhare Eslami}, \bibinfo{person}{Kristen
  Vaccaro}, \bibinfo{person}{Min~Kyung Lee}, \bibinfo{person}{Amit Elazari
  Bar~On}, \bibinfo{person}{Eric Gilbert}, {and} \bibinfo{person}{Karrie
  Karahalios}.} \bibinfo{year}{2019}\natexlab{}.
\newblock \showarticletitle{User attitudes towards algorithmic opacity and
  transparency in online reviewing platforms}. In
  \bibinfo{booktitle}{\emph{Proceedings of the 2019 CHI Conference on Human
  Factors in Computing Systems}}. \bibinfo{pages}{1--14}.
\newblock


\bibitem[\protect\citeauthoryear{Eubanks}{Eubanks}{2018}]%
        {eubanks2018automating}
\bibfield{author}{\bibinfo{person}{Virginia Eubanks}.}
  \bibinfo{year}{2018}\natexlab{}.
\newblock \bibinfo{booktitle}{\emph{Automating inequality: How high-tech tools
  profile, police, and punish the poor}}.
\newblock \bibinfo{publisher}{St. Martin's Press}.
\newblock


\bibitem[\protect\citeauthoryear{Fang, Gong, and Liu}{Fang
  et~al\mbox{.}}{2020}]%
        {Fang2020Apr}
\bibfield{author}{\bibinfo{person}{Minghong Fang},
  \bibinfo{person}{Neil~Zhenqiang Gong}, {and} \bibinfo{person}{Jia Liu}.}
  \bibinfo{year}{2020}\natexlab{}.
\newblock \bibinfo{booktitle}{\emph{{Influence Function based Data Poisoning
  Attacks to Top-N Recommender Systems}}}.
\newblock \bibinfo{publisher}{Association for Computing Machinery},
  \bibinfo{address}{New York, NY, USA}.
\newblock
\showISBNx{978-1-45037023-3}
\urldef\tempurl%
\url{https://doi.org/10.1145/3366423.3380072}
\showDOI{\tempurl}


\bibitem[\protect\citeauthoryear{Figueroa, Zeng-Treitler, Kandula, and
  Ngo}{Figueroa et~al\mbox{.}}{2012}]%
        {figueroa_predicting_2012}
\bibfield{author}{\bibinfo{person}{Rosa~L Figueroa}, \bibinfo{person}{Qing
  Zeng-Treitler}, \bibinfo{person}{Sasikiran Kandula}, {and}
  \bibinfo{person}{Long~H Ngo}.} \bibinfo{year}{2012}\natexlab{}.
\newblock \showarticletitle{Predicting sample size required for classification
  performance}.
\newblock \bibinfo{journal}{\emph{BMC medical informatics and decision making}}
  \bibinfo{volume}{12}, \bibinfo{number}{1} (\bibinfo{year}{2012}),
  \bibinfo{pages}{8}.
\newblock
\urldef\tempurl%
\url{https://link.springer.com/article/10.1186/1472-6947-12-8}
\showURL{%
\tempurl}
\newblock
\shownote{Publisher: Springer.}


\bibitem[\protect\citeauthoryear{Garcia, Mavrodiev, and Schweitzer}{Garcia
  et~al\mbox{.}}{2013}]%
        {garcia2013social}
\bibfield{author}{\bibinfo{person}{David Garcia}, \bibinfo{person}{Pavlin
  Mavrodiev}, {and} \bibinfo{person}{Frank Schweitzer}.}
  \bibinfo{year}{2013}\natexlab{}.
\newblock \showarticletitle{Social resilience in online communities: The
  autopsy of friendster}. In \bibinfo{booktitle}{\emph{Proceedings of the first
  ACM conference on Online social networks}}. \bibinfo{pages}{39--50}.
\newblock


\bibitem[\protect\citeauthoryear{Garcia, Sutherland, Cifor, Chan, Klein,
  D'Ignazio, and Salehi}{Garcia et~al\mbox{.}}{2020}]%
        {garcia2020no}
\bibfield{author}{\bibinfo{person}{Patricia Garcia}, \bibinfo{person}{Tonia
  Sutherland}, \bibinfo{person}{Marika Cifor}, \bibinfo{person}{Anita~Say
  Chan}, \bibinfo{person}{Lauren Klein}, \bibinfo{person}{Catherine D'Ignazio},
  {and} \bibinfo{person}{Niloufar Salehi}.} \bibinfo{year}{2020}\natexlab{}.
\newblock \showarticletitle{No: Critical Refusal as Feminist Data Practice}. In
  \bibinfo{booktitle}{\emph{Conference Companion Publication of the 2020 on
  Computer Supported Cooperative Work and Social Computing}}.
  \bibinfo{pages}{199--202}.
\newblock


\bibitem[\protect\citeauthoryear{Gebru}{Gebru}{2019}]%
        {gebru2019oxford}
\bibfield{author}{\bibinfo{person}{Timnit Gebru}.}
  \bibinfo{year}{2019}\natexlab{}.
\newblock \showarticletitle{Oxford Handbook on AI Ethics Book Chapter on Race
  and Gender}.
\newblock \bibinfo{journal}{\emph{arXiv preprint arXiv:1908.06165}}
  (\bibinfo{year}{2019}).
\newblock


\bibitem[\protect\citeauthoryear{Geiping, Fowl, Huang, Czaja, Taylor, Moeller,
  and Goldstein}{Geiping et~al\mbox{.}}{2020}]%
        {geiping2020witches}
\bibfield{author}{\bibinfo{person}{Jonas Geiping}, \bibinfo{person}{Liam Fowl},
  \bibinfo{person}{W.~Ronny Huang}, \bibinfo{person}{Wojciech Czaja},
  \bibinfo{person}{Gavin Taylor}, \bibinfo{person}{Michael Moeller}, {and}
  \bibinfo{person}{Tom Goldstein}.} \bibinfo{year}{2020}\natexlab{}.
\newblock \bibinfo{title}{Witches' Brew: Industrial Scale Data Poisoning via
  Gradient Matching}.
\newblock
\newblock
\showeprint[arxiv]{cs.CV/2009.02276}


\bibitem[\protect\citeauthoryear{Gillespie}{Gillespie}{2017}]%
        {gillespie2017algorithmically}
\bibfield{author}{\bibinfo{person}{Tarleton Gillespie}.}
  \bibinfo{year}{2017}\natexlab{}.
\newblock \showarticletitle{Algorithmically recognizable: Santorum’s Google
  problem, and Google’s Santorum problem}.
\newblock \bibinfo{journal}{\emph{Information, communication \& society}}
  \bibinfo{volume}{20}, \bibinfo{number}{1} (\bibinfo{year}{2017}),
  \bibinfo{pages}{63--80}.
\newblock


\bibitem[\protect\citeauthoryear{Gillespie}{Gillespie}{2018}]%
        {gillespie2018custodians}
\bibfield{author}{\bibinfo{person}{Tarleton Gillespie}.}
  \bibinfo{year}{2018}\natexlab{}.
\newblock \bibinfo{booktitle}{\emph{Custodians of the Internet: Platforms,
  content moderation, and the hidden decisions that shape social media}}.
\newblock \bibinfo{publisher}{Yale University Press}.
\newblock


\bibitem[\protect\citeauthoryear{Golebiewski and Boyd}{Golebiewski and
  Boyd}{2019}]%
        {golebiewski2019data}
\bibfield{author}{\bibinfo{person}{Michael Golebiewski} {and}
  \bibinfo{person}{Danah Boyd}.} \bibinfo{year}{2019}\natexlab{}.
\newblock \showarticletitle{Data voids: Where missing data can easily be
  exploited}.
\newblock \bibinfo{journal}{\emph{Data \& Society}} (\bibinfo{year}{2019}).
\newblock


\bibitem[\protect\citeauthoryear{Green}{Green}{2018}]%
        {green_fair_2018}
\bibfield{author}{\bibinfo{person}{Ben Green}.}
  \bibinfo{year}{2018}\natexlab{}.
\newblock \showarticletitle{‘{Fair}’ {Risk} {Assessments}: {A} {Precarious}
  {Approach} for {Criminal} {Justice} {Reform}}. In
  \bibinfo{booktitle}{\emph{5th {Workshop} on fairness, accountability, and
  transparency in machine learning}}.
\newblock


\bibitem[\protect\citeauthoryear{Greenfield, Frier, and Brody}{Greenfield
  et~al\mbox{.}}{2018}]%
        {greenfield_naacp_2018}
\bibfield{author}{\bibinfo{person}{Rebecca Greenfield}, \bibinfo{person}{Sarah
  Frier}, {and} \bibinfo{person}{Ben Brody}.} \bibinfo{year}{2018}\natexlab{}.
\newblock \showarticletitle{{NAACP} {Seeks} {Week}-{Long} {Facebook} {Boycott}
  {Over} {Racial} {Targeting}}.
\newblock \bibinfo{journal}{\emph{Bloomberg.com}} (\bibinfo{date}{Dec.}
  \bibinfo{year}{2018}).
\newblock
\urldef\tempurl%
\url{https://www.bloomberg.com/news/articles/2018-12-17/naacp-calls-for-week-long-facebook-boycott-over-racial-targeting}
\showURL{%
\tempurl}


\bibitem[\protect\citeauthoryear{Gunes, Kaleli, Bilge, and Polat}{Gunes
  et~al\mbox{.}}{2014}]%
        {gunes_shilling_2014}
\bibfield{author}{\bibinfo{person}{Ihsan Gunes}, \bibinfo{person}{Cihan
  Kaleli}, \bibinfo{person}{Alper Bilge}, {and} \bibinfo{person}{Huseyin
  Polat}.} \bibinfo{year}{2014}\natexlab{}.
\newblock \showarticletitle{Shilling attacks against recommender systems: a
  comprehensive survey}.
\newblock \bibinfo{journal}{\emph{Artificial Intelligence Review}}
  \bibinfo{volume}{42}, \bibinfo{number}{4} (\bibinfo{year}{2014}),
  \bibinfo{pages}{767--799}.
\newblock
\newblock
\shownote{Publisher: Springer.}


\bibitem[\protect\citeauthoryear{Gurstein}{Gurstein}{2011}]%
        {Gurstein}
\bibfield{author}{\bibinfo{person}{Michael Gurstein}.}
  \bibinfo{year}{2011}\natexlab{}.
\newblock \showarticletitle{Open data: Empowering the empowered or effective
  data use for everyone?}
\newblock \bibinfo{journal}{\emph{First Monday}}  \bibinfo{volume}{16}
  (\bibinfo{date}{02} \bibinfo{year}{2011}).
\newblock
\urldef\tempurl%
\url{https://doi.org/10.5210/fm.v16i2.3316}
\showDOI{\tempurl}


\bibitem[\protect\citeauthoryear{Hancock, Toma, and Ellison}{Hancock
  et~al\mbox{.}}{2007}]%
        {hancock2007truth}
\bibfield{author}{\bibinfo{person}{Jeffrey~T Hancock},
  \bibinfo{person}{Catalina Toma}, {and} \bibinfo{person}{Nicole Ellison}.}
  \bibinfo{year}{2007}\natexlab{}.
\newblock \showarticletitle{The truth about lying in online dating profiles}.
  In \bibinfo{booktitle}{\emph{Proceedings of the SIGCHI conference on Human
  factors in computing systems}}. \bibinfo{pages}{449--452}.
\newblock


\bibitem[\protect\citeauthoryear{Hara, Adams, Milland, Savage, Callison-Burch,
  and Bigham}{Hara et~al\mbox{.}}{2018}]%
        {hara_data-driven_2018}
\bibfield{author}{\bibinfo{person}{Kotaro Hara}, \bibinfo{person}{Abigail
  Adams}, \bibinfo{person}{Kristy Milland}, \bibinfo{person}{Saiph Savage},
  \bibinfo{person}{Chris Callison-Burch}, {and} \bibinfo{person}{Jeffrey~P
  Bigham}.} \bibinfo{year}{2018}\natexlab{}.
\newblock \showarticletitle{A {Data}-{Driven} {Analysis} of {Workers}'
  {Earnings} on {Amazon} {Mechanical} {Turk}}. In
  \bibinfo{booktitle}{\emph{Proceedings of the 2018 {CHI} {Conference} on
  {Human} {Factors} in {Computing} {Systems}}}. \bibinfo{publisher}{ACM},
  \bibinfo{pages}{449}.
\newblock


\bibitem[\protect\citeauthoryear{Hecht, Wilcox, Bigham, Schöning, Hoque,
  Ernst, Bisk, De~Russis, Yarosh, Anjum, and {others}}{Hecht
  et~al\mbox{.}}{2018}]%
        {hecht_its_2018}
\bibfield{author}{\bibinfo{person}{B Hecht}, \bibinfo{person}{L Wilcox},
  \bibinfo{person}{JP Bigham}, \bibinfo{person}{J Schöning},
  \bibinfo{person}{E Hoque}, \bibinfo{person}{J Ernst}, \bibinfo{person}{Y
  Bisk}, \bibinfo{person}{L De~Russis}, \bibinfo{person}{L Yarosh},
  \bibinfo{person}{B Anjum}, {and} \bibinfo{person}{{others}}.}
  \bibinfo{year}{2018}\natexlab{}.
\newblock \bibinfo{title}{It's time to do something: {Mitigating} the negative
  impacts of computing through a change to the peer review process.}
\newblock
\newblock


\bibitem[\protect\citeauthoryear{Hestness, Narang, Ardalani, Diamos, Jun,
  Kianinejad, Patwary, Ali, Yang, and Zhou}{Hestness et~al\mbox{.}}{2017}]%
        {hestness2017deep}
\bibfield{author}{\bibinfo{person}{Joel Hestness}, \bibinfo{person}{Sharan
  Narang}, \bibinfo{person}{Newsha Ardalani}, \bibinfo{person}{Gregory Diamos},
  \bibinfo{person}{Heewoo Jun}, \bibinfo{person}{Hassan Kianinejad},
  \bibinfo{person}{Md Patwary}, \bibinfo{person}{Mostofa Ali},
  \bibinfo{person}{Yang Yang}, {and} \bibinfo{person}{Yanqi Zhou}.}
  \bibinfo{year}{2017}\natexlab{}.
\newblock \showarticletitle{Deep learning scaling is predictable, empirically}.
\newblock \bibinfo{journal}{\emph{arXiv preprint arXiv:1712.00409}}
  (\bibinfo{year}{2017}).
\newblock


\bibitem[\protect\citeauthoryear{Hinnosaar, Hinnosaar, Kummer, and
  Slivko}{Hinnosaar et~al\mbox{.}}{2019}]%
        {hinnosaar2019wikipedia}
\bibfield{author}{\bibinfo{person}{Marit Hinnosaar}, \bibinfo{person}{Toomas
  Hinnosaar}, \bibinfo{person}{Michael~E Kummer}, {and} \bibinfo{person}{Olga
  Slivko}.} \bibinfo{year}{2019}\natexlab{}.
\newblock \showarticletitle{Wikipedia matters}.
\newblock \bibinfo{journal}{\emph{Available at SSRN 3046400}}
  (\bibinfo{year}{2019}).
\newblock


\bibitem[\protect\citeauthoryear{Howe and Nissenbaum}{Howe and
  Nissenbaum}{2017}]%
        {howe2017engineering}
\bibfield{author}{\bibinfo{person}{Daniel~C Howe} {and} \bibinfo{person}{Helen
  Nissenbaum}.} \bibinfo{year}{2017}\natexlab{}.
\newblock \showarticletitle{Engineering Privacy and Protest: A Case Study of
  AdNauseam.}. In \bibinfo{booktitle}{\emph{IWPE@ SP}}.
  \bibinfo{pages}{57--64}.
\newblock


\bibitem[\protect\citeauthoryear{Hunt}{Hunt}{2015}]%
        {hunt2015gaming}
\bibfield{author}{\bibinfo{person}{Kate~Mathews Hunt}.}
  \bibinfo{year}{2015}\natexlab{}.
\newblock \showarticletitle{Gaming the system: Fake online reviews v. consumer
  law}.
\newblock \bibinfo{journal}{\emph{Computer law \& security review}}
  \bibinfo{volume}{31}, \bibinfo{number}{1} (\bibinfo{year}{2015}),
  \bibinfo{pages}{3--25}.
\newblock


\bibitem[\protect\citeauthoryear{Jackson, Bailey, and Welles}{Jackson
  et~al\mbox{.}}{2020}]%
        {jackson2020hashtagactivism}
\bibfield{author}{\bibinfo{person}{Sarah~J Jackson}, \bibinfo{person}{Moya
  Bailey}, {and} \bibinfo{person}{Brooke~Foucault Welles}.}
  \bibinfo{year}{2020}\natexlab{}.
\newblock \bibinfo{booktitle}{\emph{\# HashtagActivism: Networks of Race and
  Gender Justice}}.
\newblock \bibinfo{publisher}{MIT Press}.
\newblock


\bibitem[\protect\citeauthoryear{James}{James}{2020}]%
        {James2020Jan}
\bibfield{author}{\bibinfo{person}{Ross James}.}
  \bibinfo{year}{2020}\natexlab{}.
\newblock \showarticletitle{{How to use Google Takeout to download your Google
  data - Business Insider}}.
\newblock \bibinfo{journal}{\emph{Business Insider}} (\bibinfo{date}{Jan}
  \bibinfo{year}{2020}).
\newblock
\urldef\tempurl%
\url{https://www.businessinsider.com/what-is-google-takeout}
\showURL{%
\tempurl}


\bibitem[\protect\citeauthoryear{Jia, Dao, Wang, Hubis, Hynes, G{\"u}rel, Li,
  Zhang, Song, and Spanos}{Jia et~al\mbox{.}}{2019}]%
        {jia2019towards}
\bibfield{author}{\bibinfo{person}{Ruoxi Jia}, \bibinfo{person}{David Dao},
  \bibinfo{person}{Boxin Wang}, \bibinfo{person}{Frances~Ann Hubis},
  \bibinfo{person}{Nick Hynes}, \bibinfo{person}{Nezihe~Merve G{\"u}rel},
  \bibinfo{person}{Bo Li}, \bibinfo{person}{Ce Zhang}, \bibinfo{person}{Dawn
  Song}, {and} \bibinfo{person}{Costas~J Spanos}.}
  \bibinfo{year}{2019}\natexlab{}.
\newblock \showarticletitle{Towards Efficient Data Valuation Based on the
  Shapley Value}. In \bibinfo{booktitle}{\emph{The 22nd International
  Conference on Artificial Intelligence and Statistics}}.
  \bibinfo{pages}{1167--1176}.
\newblock


\bibitem[\protect\citeauthoryear{Johnson, Lin, Li, Hall, Halfaker,
  Sch\"{o}ning, and Hecht}{Johnson et~al\mbox{.}}{2016}]%
        {10.1145/2858036.2858123}
\bibfield{author}{\bibinfo{person}{Isaac~L. Johnson}, \bibinfo{person}{Yilun
  Lin}, \bibinfo{person}{Toby Jia-Jun Li}, \bibinfo{person}{Andrew Hall},
  \bibinfo{person}{Aaron Halfaker}, \bibinfo{person}{Johannes Sch\"{o}ning},
  {and} \bibinfo{person}{Brent Hecht}.} \bibinfo{year}{2016}\natexlab{}.
\newblock \bibinfo{booktitle}{\emph{Not at Home on the Range: Peer Production
  and the Urban/Rural Divide}}.
\newblock \bibinfo{publisher}{Association for Computing Machinery},
  \bibinfo{address}{New York, NY, USA}, \bibinfo{pages}{13–25}.
\newblock
\showISBNx{9781450333627}
\urldef\tempurl%
\url{https://doi.org/10.1145/2858036.2858123}
\showURL{%
\tempurl}


\bibitem[\protect\citeauthoryear{Jones and Tonetti}{Jones and Tonetti}{2019}]%
        {jones_nonrivalry_2019}
\bibfield{author}{\bibinfo{person}{Charles~I Jones} {and}
  \bibinfo{person}{Christopher Tonetti}.} \bibinfo{year}{2019}\natexlab{}.
\newblock \bibinfo{booktitle}{\emph{Nonrivalry and the {Economics} of {Data}}}.
\newblock \bibinfo{type}{{T}echnical {R}eport}. \bibinfo{institution}{National
  Bureau of Economic Research}.
\newblock


\bibitem[\protect\citeauthoryear{Koh, Ang, Teo, and Liang}{Koh
  et~al\mbox{.}}{2019}]%
        {koh2019accuracy}
\bibfield{author}{\bibinfo{person}{Pang Wei~W Koh}, \bibinfo{person}{Kai-Siang
  Ang}, \bibinfo{person}{Hubert Teo}, {and} \bibinfo{person}{Percy~S Liang}.}
  \bibinfo{year}{2019}\natexlab{}.
\newblock \showarticletitle{On the accuracy of influence functions for
  measuring group effects}. In \bibinfo{booktitle}{\emph{Advances in Neural
  Information Processing Systems}}. \bibinfo{pages}{5254--5264}.
\newblock


\bibitem[\protect\citeauthoryear{Koos}{Koos}{2012}]%
        {koos_what_2012}
\bibfield{author}{\bibinfo{person}{Sebastian Koos}.}
  \bibinfo{year}{2012}\natexlab{}.
\newblock \showarticletitle{What drives political consumption in {Europe}? {A}
  multi-level analysis on individual characteristics, opportunity structures
  and globalization}.
\newblock \bibinfo{journal}{\emph{Acta Sociologica}} \bibinfo{volume}{55},
  \bibinfo{number}{1} (\bibinfo{date}{March} \bibinfo{year}{2012}),
  \bibinfo{pages}{37--57}.
\newblock
\showISSN{0001-6993}
\urldef\tempurl%
\url{https://doi.org/10.1177/0001699311431594}
\showDOI{\tempurl}


\bibitem[\protect\citeauthoryear{Kraut, Resnick, Kiesler, Burke, Chen, Kittur,
  Konstan, Ren, and Riedl}{Kraut et~al\mbox{.}}{2012}]%
        {kraut_building_2012}
\bibfield{author}{\bibinfo{person}{Robert~E Kraut}, \bibinfo{person}{Paul
  Resnick}, \bibinfo{person}{Sara Kiesler}, \bibinfo{person}{Moira Burke},
  \bibinfo{person}{Yan Chen}, \bibinfo{person}{Niki Kittur},
  \bibinfo{person}{Joseph Konstan}, \bibinfo{person}{Yuqing Ren}, {and}
  \bibinfo{person}{John Riedl}.} \bibinfo{year}{2012}\natexlab{}.
\newblock \bibinfo{booktitle}{\emph{Building successful online communities:
  {Evidence}-based social design}}.
\newblock \bibinfo{publisher}{Mit Press}.
\newblock


\bibitem[\protect\citeauthoryear{Kulynych, Overdorf, Troncoso, and
  Gürses}{Kulynych et~al\mbox{.}}{2020}]%
        {kulynych_pots_2020}
\bibfield{author}{\bibinfo{person}{Bogdan Kulynych}, \bibinfo{person}{Rebekah
  Overdorf}, \bibinfo{person}{Carmela Troncoso}, {and} \bibinfo{person}{Seda
  Gürses}.} \bibinfo{year}{2020}\natexlab{}.
\newblock \showarticletitle{{POTs}: protective optimization technologies}. In
  \bibinfo{booktitle}{\emph{Proceedings of the 2020 {Conference} on {Fairness},
  {Accountability}, and {Transparency}}}. \bibinfo{pages}{177--188}.
\newblock


\bibitem[\protect\citeauthoryear{Kumar, Zafarani, and Liu}{Kumar
  et~al\mbox{.}}{2011}]%
        {kumar2011understanding}
\bibfield{author}{\bibinfo{person}{Shamanth Kumar}, \bibinfo{person}{Reza
  Zafarani}, {and} \bibinfo{person}{Huan Liu}.}
  \bibinfo{year}{2011}\natexlab{}.
\newblock \showarticletitle{Understanding User Migration Patterns in Social
  Media.}. In \bibinfo{booktitle}{\emph{AAAI}}, Vol.~\bibinfo{volume}{11}.
  \bibinfo{pages}{8--11}.
\newblock


\bibitem[\protect\citeauthoryear{Lam and Riedl}{Lam and Riedl}{2004}]%
        {lam2004shilling}
\bibfield{author}{\bibinfo{person}{Shyong~K Lam} {and} \bibinfo{person}{John
  Riedl}.} \bibinfo{year}{2004}\natexlab{}.
\newblock \showarticletitle{Shilling recommender systems for fun and profit}.
  In \bibinfo{booktitle}{\emph{Proceedings of the 13th international conference
  on World Wide Web}}. \bibinfo{pages}{393--402}.
\newblock


\bibitem[\protect\citeauthoryear{Lampe, Ellison, and Steinfeld}{Lampe
  et~al\mbox{.}}{2008}]%
        {lampe_use_perception}
\bibfield{author}{\bibinfo{person}{Cliff Lampe}, \bibinfo{person}{Nicole~B.
  Ellison}, {and} \bibinfo{person}{Charles Steinfeld}.}
  \bibinfo{year}{2008}\natexlab{}.
\newblock \showarticletitle{Changes in Use and Perception of Facebook}. In
  \bibinfo{booktitle}{\emph{Proceedings of the 2008 conference on {Computer}
  supported cooperative work}}. \bibinfo{pages}{721--730}.
\newblock


\bibitem[\protect\citeauthoryear{Lampe, Vitak, and Ellison}{Lampe
  et~al\mbox{.}}{2013}]%
        {lampe_users_2013}
\bibfield{author}{\bibinfo{person}{Cliff Lampe}, \bibinfo{person}{Jessica
  Vitak}, {and} \bibinfo{person}{Nicole Ellison}.}
  \bibinfo{year}{2013}\natexlab{}.
\newblock \showarticletitle{Users and nonusers: {Interactions} between levels
  of adoption and social capital}. In \bibinfo{booktitle}{\emph{Proceedings of
  the 2013 conference on {Computer} supported cooperative work}}.
  \bibinfo{pages}{809--820}.
\newblock


\bibitem[\protect\citeauthoryear{Lee and Zhu}{Lee and Zhu}{2012}]%
        {lee_shilling_2012}
\bibfield{author}{\bibinfo{person}{Jong-Seok Lee} {and} \bibinfo{person}{Dan
  Zhu}.} \bibinfo{year}{2012}\natexlab{}.
\newblock \showarticletitle{Shilling attack detection—a new approach for a
  trustworthy recommender system}.
\newblock \bibinfo{journal}{\emph{INFORMS Journal on Computing}}
  \bibinfo{volume}{24}, \bibinfo{number}{1} (\bibinfo{year}{2012}),
  \bibinfo{pages}{117--131}.
\newblock
\newblock
\shownote{Publisher: INFORMS.}


\bibitem[\protect\citeauthoryear{Lee, Kusbit, Metsky, and Dabbish}{Lee
  et~al\mbox{.}}{2015}]%
        {lee2015working}
\bibfield{author}{\bibinfo{person}{Min~Kyung Lee}, \bibinfo{person}{Daniel
  Kusbit}, \bibinfo{person}{Evan Metsky}, {and} \bibinfo{person}{Laura
  Dabbish}.} \bibinfo{year}{2015}\natexlab{}.
\newblock \showarticletitle{Working with machines: The impact of algorithmic
  and data-driven management on human workers}. In
  \bibinfo{booktitle}{\emph{Proceedings of the 33rd annual ACM conference on
  human factors in computing systems}}. \bibinfo{pages}{1603--1612}.
\newblock


\bibitem[\protect\citeauthoryear{Lehtiniemi and Ruckenstein}{Lehtiniemi and
  Ruckenstein}{2018}]%
        {lehtiniemi2018social}
\bibfield{author}{\bibinfo{person}{Tuukka Lehtiniemi} {and}
  \bibinfo{person}{Minna Ruckenstein}.} \bibinfo{year}{2018}\natexlab{}.
\newblock \showarticletitle{The social imaginaries of data activism}.
\newblock \bibinfo{journal}{\emph{Big Data \& Society}} \bibinfo{volume}{6},
  \bibinfo{number}{1} (\bibinfo{year}{2018}),
  \bibinfo{pages}{2053951718821146}.
\newblock


\bibitem[\protect\citeauthoryear{Li, Wang, Singh, and Vorobeychik}{Li
  et~al\mbox{.}}{2016}]%
        {li2016data}
\bibfield{author}{\bibinfo{person}{Bo Li}, \bibinfo{person}{Yining Wang},
  \bibinfo{person}{Aarti Singh}, {and} \bibinfo{person}{Yevgeniy Vorobeychik}.}
  \bibinfo{year}{2016}\natexlab{}.
\newblock \showarticletitle{Data poisoning attacks on factorization-based
  collaborative filtering}. In \bibinfo{booktitle}{\emph{Advances in neural
  information processing systems}}. \bibinfo{pages}{1885--1893}.
\newblock


\bibitem[\protect\citeauthoryear{Li, Alarcon, Espinosa, and Hecht}{Li
  et~al\mbox{.}}{2018}]%
        {li_out_2018}
\bibfield{author}{\bibinfo{person}{Hanlin Li}, \bibinfo{person}{Bodhi Alarcon},
  \bibinfo{person}{Sara~M. Espinosa}, {and} \bibinfo{person}{Brent Hecht}.}
  \bibinfo{year}{2018}\natexlab{}.
\newblock \showarticletitle{Out of {Site}: {Empowering} a {New} {Approach} to
  {Online} {Boycotts}}.
\newblock \bibinfo{journal}{\emph{Proceedings of the 2018 Computer-Supported
  Cooperative Work and Social Computing (CSCW’2018 / PACM)}}
  (\bibinfo{year}{2018}).
\newblock


\bibitem[\protect\citeauthoryear{Li and Hecht}{Li and Hecht}{2020}]%
        {li_rating_2020}
\bibfield{author}{\bibinfo{person}{Hanlin Li} {and} \bibinfo{person}{Brent
  Hecht}.} \bibinfo{year}{2020}\natexlab{}.
\newblock \showarticletitle{3 Stars on Yelp, 4 Stars on Google Maps: A
  Cross-Platform Examination of Restaurant Ratings}.
\newblock \bibinfo{journal}{\emph{Proceedings of the ACM on Human-Computer
  Interaction}} \bibinfo{volume}{4}, \bibinfo{number}{CSCW}
  (\bibinfo{year}{2020}).
\newblock


\bibitem[\protect\citeauthoryear{Li, Vincent, Tsai, Kaye, and Hecht}{Li
  et~al\mbox{.}}{2019}]%
        {li_how_2019}
\bibfield{author}{\bibinfo{person}{Hanlin Li}, \bibinfo{person}{Nicholas
  Vincent}, \bibinfo{person}{Janice Tsai}, \bibinfo{person}{Jofish Kaye}, {and}
  \bibinfo{person}{Brent Hecht}.} \bibinfo{year}{2019}\natexlab{}.
\newblock \showarticletitle{How do people change their technology use in
  protest?: {Understanding} “protest users”}.
\newblock \bibinfo{journal}{\emph{Proceedings of the ACM on Human-Computer
  Interaction}} \bibinfo{volume}{3}, \bibinfo{number}{CSCW}
  (\bibinfo{year}{2019}), \bibinfo{pages}{87}.
\newblock


\bibitem[\protect\citeauthoryear{Li, Ott, Cardie, and Hovy}{Li
  et~al\mbox{.}}{2014}]%
        {li2014towards}
\bibfield{author}{\bibinfo{person}{Jiwei Li}, \bibinfo{person}{Myle Ott},
  \bibinfo{person}{Claire Cardie}, {and} \bibinfo{person}{Eduard Hovy}.}
  \bibinfo{year}{2014}\natexlab{}.
\newblock \showarticletitle{Towards a general rule for identifying deceptive
  opinion spam}. In \bibinfo{booktitle}{\emph{Proceedings of the 52nd Annual
  Meeting of the Association for Computational Linguistics (Volume 1: Long
  Papers)}}. \bibinfo{pages}{1566--1576}.
\newblock


\bibitem[\protect\citeauthoryear{Lyons}{Lyons}{2021}]%
        {lyons_ftc_2021}
\bibfield{author}{\bibinfo{person}{Kim Lyons}.}
  \bibinfo{year}{2021}\natexlab{}.
\newblock \showarticletitle{{FTC} settles with photo storage app that pivoted
  to facial recognition}.
\newblock \bibinfo{journal}{\emph{The Verge}} (\bibinfo{date}{Jan.}
  \bibinfo{year}{2021}).
\newblock
\urldef\tempurl%
\url{https://www.theverge.com/2021/1/11/22225171/ftc-facial-recognition-ever-settled-paravision-privacy-photos}
\showURL{%
\tempurl}
\newblock
\shownote{Publisher: The Verge.}


\bibitem[\protect\citeauthoryear{Margetts, John, Hale, and Yasseri}{Margetts
  et~al\mbox{.}}{2015}]%
        {margetts2015political}
\bibfield{author}{\bibinfo{person}{Helen Margetts}, \bibinfo{person}{Peter
  John}, \bibinfo{person}{Scott Hale}, {and} \bibinfo{person}{Taha Yasseri}.}
  \bibinfo{year}{2015}\natexlab{}.
\newblock \bibinfo{booktitle}{\emph{Political turbulence: How social media
  shape collective action}}.
\newblock \bibinfo{publisher}{Princeton University Press}.
\newblock


\bibitem[\protect\citeauthoryear{Mathur, Vitak, Narayanan, and Chetty}{Mathur
  et~al\mbox{.}}{2018}]%
        {mathur2018characterizing}
\bibfield{author}{\bibinfo{person}{Arunesh Mathur}, \bibinfo{person}{Jessica
  Vitak}, \bibinfo{person}{Arvind Narayanan}, {and} \bibinfo{person}{Marshini
  Chetty}.} \bibinfo{year}{2018}\natexlab{}.
\newblock \showarticletitle{Characterizing the use of browser-based blocking
  extensions to prevent online tracking}. In
  \bibinfo{booktitle}{\emph{Fourteenth Symposium on Usable Privacy and Security
  ($\{$SOUPS$\}$ 2018)}}. \bibinfo{pages}{103--116}.
\newblock


\bibitem[\protect\citeauthoryear{Matias}{Matias}{2016}]%
        {matias2016going}
\bibfield{author}{\bibinfo{person}{J~Nathan Matias}.}
  \bibinfo{year}{2016}\natexlab{}.
\newblock \showarticletitle{Going dark: Social factors in collective action
  against platform operators in the Reddit blackout}. In
  \bibinfo{booktitle}{\emph{Proceedings of the 2016 CHI conference on human
  factors in computing systems}}. \bibinfo{pages}{1138--1151}.
\newblock


\bibitem[\protect\citeauthoryear{McMahon, Johnson, and Hecht}{McMahon
  et~al\mbox{.}}{2017}]%
        {mcmahon_substantial_2017}
\bibfield{author}{\bibinfo{person}{Connor McMahon}, \bibinfo{person}{Isaac~L
  Johnson}, {and} \bibinfo{person}{Brent Hecht}.}
  \bibinfo{year}{2017}\natexlab{}.
\newblock \showarticletitle{The {Substantial} {Interdependence} of {Wikipedia}
  and {Google}: {A} {Case} {Study} on the {Relationship} {Between} {Peer}
  {Production} {Communities} and {Information} {Technologies}.}. In
  \bibinfo{booktitle}{\emph{{ICWSM}}}. \bibinfo{pages}{142--151}.
\newblock


\bibitem[\protect\citeauthoryear{Milan and Van~der Velden}{Milan and Van~der
  Velden}{2016}]%
        {milan2016alternative}
\bibfield{author}{\bibinfo{person}{Stefania Milan} {and}
  \bibinfo{person}{Lonneke Van~der Velden}.} \bibinfo{year}{2016}\natexlab{}.
\newblock \showarticletitle{The alternative epistemologies of data activism}.
\newblock \bibinfo{journal}{\emph{Digital Culture \& Society}}
  \bibinfo{volume}{2}, \bibinfo{number}{2} (\bibinfo{year}{2016}),
  \bibinfo{pages}{57--74}.
\newblock


\bibitem[\protect\citeauthoryear{Mobasher, Burke, Bhaumik, and
  Williams}{Mobasher et~al\mbox{.}}{2005}]%
        {mobasher2005effective}
\bibfield{author}{\bibinfo{person}{Bamshad Mobasher}, \bibinfo{person}{Robin
  Burke}, \bibinfo{person}{Runa Bhaumik}, {and} \bibinfo{person}{Chad
  Williams}.} \bibinfo{year}{2005}\natexlab{}.
\newblock \showarticletitle{Effective attack models for shilling item-based
  collaborative filtering systems}. In \bibinfo{booktitle}{\emph{Proceedings of
  the WebKDD Workshop}}. Citeseer, \bibinfo{pages}{13--23}.
\newblock


\bibitem[\protect\citeauthoryear{Nechaev, Corcoglioniti, and Giuliano}{Nechaev
  et~al\mbox{.}}{2017}]%
        {nechaev_concealing_2017}
\bibfield{author}{\bibinfo{person}{Yaroslav Nechaev},
  \bibinfo{person}{Francesco Corcoglioniti}, {and} \bibinfo{person}{Claudio
  Giuliano}.} \bibinfo{year}{2017}\natexlab{}.
\newblock \showarticletitle{Concealing {Interests} of {Passive} {Users} in
  {Social} {Media}.}. In \bibinfo{booktitle}{\emph{{BlackMirror}@ {ISWC}}}.
\newblock


\bibitem[\protect\citeauthoryear{Newell, Jurgens, Saleem, Vala, Sassine,
  Armstrong, and Ruths}{Newell et~al\mbox{.}}{2016}]%
        {newell2016user}
\bibfield{author}{\bibinfo{person}{Edward Newell}, \bibinfo{person}{David
  Jurgens}, \bibinfo{person}{Haji~Mohammad Saleem}, \bibinfo{person}{Hardik
  Vala}, \bibinfo{person}{Jad Sassine}, \bibinfo{person}{Caitrin Armstrong},
  {and} \bibinfo{person}{Derek Ruths}.} \bibinfo{year}{2016}\natexlab{}.
\newblock \showarticletitle{User Migration in Online Social Networks: A Case
  Study on Reddit During a Period of Community Unrest.}. In
  \bibinfo{booktitle}{\emph{ICWSM}}. \bibinfo{pages}{279--288}.
\newblock


\bibitem[\protect\citeauthoryear{Noble}{Noble}{2018}]%
        {noble2018algorithms}
\bibfield{author}{\bibinfo{person}{Safiya~Umoja Noble}.}
  \bibinfo{year}{2018}\natexlab{}.
\newblock \bibinfo{booktitle}{\emph{Algorithms of oppression: How search
  engines reinforce racism}}.
\newblock \bibinfo{publisher}{nyu Press}.
\newblock


\bibitem[\protect\citeauthoryear{Ott, Cardie, and Hancock}{Ott
  et~al\mbox{.}}{2012}]%
        {ott2012estimating}
\bibfield{author}{\bibinfo{person}{Myle Ott}, \bibinfo{person}{Claire Cardie},
  {and} \bibinfo{person}{Jeff Hancock}.} \bibinfo{year}{2012}\natexlab{}.
\newblock \showarticletitle{Estimating the prevalence of deception in online
  review communities}. In \bibinfo{booktitle}{\emph{Proceedings of the 21st
  international conference on World Wide Web}}. \bibinfo{pages}{201--210}.
\newblock


\bibitem[\protect\citeauthoryear{Paul}{Paul}{2020}]%
        {Paul2020Apr}
\bibfield{author}{\bibinfo{person}{Kari Paul}.}
  \bibinfo{year}{2020}\natexlab{}.
\newblock \showarticletitle{{Prime Day: activists protest against Amazon in
  cities across US}}.
\newblock \bibinfo{journal}{\emph{the Guardian}} (\bibinfo{date}{Apr}
  \bibinfo{year}{2020}).
\newblock
\urldef\tempurl%
\url{https://www.theguardian.com/technology/2019/jul/15/prime-day-activists-plan-protests-in-us-cities-and-a-boycott-of-e-commerce-giant}
\showURL{%
\tempurl}


\bibitem[\protect\citeauthoryear{Paul R. La~Monica}{Paul R. La~Monica}{2020}]%
        {PaulR.LaMonica2020Oct}
\bibfield{author}{\bibinfo{person}{Business Paul R. La~Monica}.}
  \bibinfo{year}{2020}\natexlab{}.
\newblock \bibinfo{title}{{Tech's magnificent seven are worth
  {\ifmmode\$\else\textdollar\fi}7.7 trillion}}.
\newblock
\newblock
\urldef\tempurl%
\url{https://www.cnn.com/2020/08/20/investing/faang-microsoft-tesla/index.html}
\showURL{%
\tempurl}
\newblock
\shownote{[Online; accessed 6. Oct. 2020].}


\bibitem[\protect\citeauthoryear{Pitropakis, Panaousis, Giannetsos,
  Anastasiadis, and Loukas}{Pitropakis et~al\mbox{.}}{2019}]%
        {pitropakis_taxonomy_2019}
\bibfield{author}{\bibinfo{person}{Nikolaos Pitropakis},
  \bibinfo{person}{Emmanouil Panaousis}, \bibinfo{person}{Thanassis
  Giannetsos}, \bibinfo{person}{Eleftherios Anastasiadis}, {and}
  \bibinfo{person}{George Loukas}.} \bibinfo{year}{2019}\natexlab{}.
\newblock \showarticletitle{A taxonomy and survey of attacks against machine
  learning}.
\newblock \bibinfo{journal}{\emph{Computer Science Review}}
  \bibinfo{volume}{34} (\bibinfo{date}{Nov.} \bibinfo{year}{2019}),
  \bibinfo{pages}{100199}.
\newblock
\showISSN{1574-0137}
\urldef\tempurl%
\url{https://doi.org/10.1016/j.cosrev.2019.100199}
\showDOI{\tempurl}
\newblock
\shownote{Publisher: Elsevier.}


\bibitem[\protect\citeauthoryear{Portwood-Stacer}{Portwood-Stacer}{2013}]%
        {portwood2013media}
\bibfield{author}{\bibinfo{person}{Laura Portwood-Stacer}.}
  \bibinfo{year}{2013}\natexlab{}.
\newblock \showarticletitle{Media refusal and conspicuous non-consumption: The
  performative and political dimensions of Facebook abstention}.
\newblock \bibinfo{journal}{\emph{New Media \& Society}} \bibinfo{volume}{15},
  \bibinfo{number}{7} (\bibinfo{year}{2013}), \bibinfo{pages}{1041--1057}.
\newblock


\bibitem[\protect\citeauthoryear{Posner and Weyl}{Posner and Weyl}{2018}]%
        {posner_radical_2018}
\bibfield{author}{\bibinfo{person}{Eric~A Posner} {and} \bibinfo{person}{E~Glen
  Weyl}.} \bibinfo{year}{2018}\natexlab{}.
\newblock \bibinfo{booktitle}{\emph{Radical {Markets}: {Uprooting} {Capitalism}
  and {Democracy} for a {Just} {Society}}}.
\newblock \bibinfo{publisher}{Princeton University Press}.
\newblock


\bibitem[\protect\citeauthoryear{Prainsack}{Prainsack}{2019}]%
        {prainsack_data_2019}
\bibfield{author}{\bibinfo{person}{Barbara Prainsack}.}
  \bibinfo{year}{2019}\natexlab{}.
\newblock \showarticletitle{Data donation: {How} to resist the {iLeviathan}}.
\newblock In \bibinfo{booktitle}{\emph{The ethics of medical data donation}}.
  \bibinfo{publisher}{Springer, Cham}, \bibinfo{pages}{9--22}.
\newblock


\bibitem[\protect\citeauthoryear{Quinn and Bederson}{Quinn and
  Bederson}{2011}]%
        {quinn_human_2011}
\bibfield{author}{\bibinfo{person}{Alexander~J Quinn} {and}
  \bibinfo{person}{Benjamin~B Bederson}.} \bibinfo{year}{2011}\natexlab{}.
\newblock \showarticletitle{Human computation: a survey and taxonomy of a
  growing field}. In \bibinfo{booktitle}{\emph{Proceedings of the {SIGCHI}
  conference on human factors in computing systems}}. \bibinfo{publisher}{ACM},
  \bibinfo{pages}{1403--1412}.
\newblock


\bibitem[\protect\citeauthoryear{Satariano}{Satariano}{2020}]%
        {satariano_what_2020}
\bibfield{author}{\bibinfo{person}{Adam Satariano}.}
  \bibinfo{year}{2020}\natexlab{}.
\newblock \showarticletitle{What the {G}.{D}.{P}.{R}., {Europe}'s {Tough} {New}
  {Data} {Law}, {Means} for {You}}.
\newblock \bibinfo{journal}{\emph{N.Y. Times}} (\bibinfo{date}{May}
  \bibinfo{year}{2020}).
\newblock
\showISSN{0362-4331}
\urldef\tempurl%
\url{https://www.nytimes.com/2018/05/06/technology/gdpr-european-privacy-law.html}
\showURL{%
\tempurl}
\newblock
\shownote{Publisher: The New York Times Company.}


\bibitem[\protect\citeauthoryear{Satchell and Dourish}{Satchell and
  Dourish}{2009}]%
        {satchell_beyond_2009}
\bibfield{author}{\bibinfo{person}{Christine Satchell} {and}
  \bibinfo{person}{Paul Dourish}.} \bibinfo{year}{2009}\natexlab{}.
\newblock \showarticletitle{Beyond the user: use and non-use in {HCI}}. In
  \bibinfo{booktitle}{\emph{Proceedings of the 21st {Annual} {Conference} of
  the {Australian} {Computer}-{Human} {Interaction} {Special} {Interest}
  {Group}: {Design}: {Open} 24/7}}. \bibinfo{pages}{9--16}.
\newblock


\bibitem[\protect\citeauthoryear{Saxena, Skeba, Guha, and Baumer}{Saxena
  et~al\mbox{.}}{2020}]%
        {saxena2020methods}
\bibfield{author}{\bibinfo{person}{Devansh Saxena}, \bibinfo{person}{Patrick
  Skeba}, \bibinfo{person}{Shion Guha}, {and} \bibinfo{person}{Eric~PS
  Baumer}.} \bibinfo{year}{2020}\natexlab{}.
\newblock \showarticletitle{Methods for Generating Typologies of Non/use}.
\newblock \bibinfo{journal}{\emph{Proceedings of the ACM on Human-Computer
  Interaction}} \bibinfo{volume}{4}, \bibinfo{number}{CSCW1}
  (\bibinfo{year}{2020}), \bibinfo{pages}{1--26}.
\newblock


\bibitem[\protect\citeauthoryear{Schoenebeck}{Schoenebeck}{2014}]%
        {schoenebeck_giving_2014}
\bibfield{author}{\bibinfo{person}{Sarita~Yardi Schoenebeck}.}
  \bibinfo{year}{2014}\natexlab{}.
\newblock \showarticletitle{Giving up {Twitter} for {Lent}: how and why we take
  breaks from social media}. In \bibinfo{booktitle}{\emph{Proceedings of the
  {SIGCHI} {Conference} on {Human} {Factors} in {Computing} {Systems}}}.
  \bibinfo{pages}{773--782}.
\newblock


\bibitem[\protect\citeauthoryear{Schwartz, Dodge, Smith, and Etzioni}{Schwartz
  et~al\mbox{.}}{2020}]%
        {schwartz2019green}
\bibfield{author}{\bibinfo{person}{Roy Schwartz}, \bibinfo{person}{Jesse
  Dodge}, \bibinfo{person}{Noah~A. Smith}, {and} \bibinfo{person}{Oren
  Etzioni}.} \bibinfo{year}{2020}\natexlab{}.
\newblock \showarticletitle{Green AI}.
\newblock \bibinfo{journal}{\emph{Commun. ACM}} \bibinfo{volume}{63},
  \bibinfo{number}{12} (\bibinfo{date}{Nov.} \bibinfo{year}{2020}),
  \bibinfo{pages}{54–63}.
\newblock
\showISSN{0001-0782}
\urldef\tempurl%
\url{https://doi.org/10.1145/3381831}
\showDOI{\tempurl}


\bibitem[\protect\citeauthoryear{Selwyn}{Selwyn}{2003}]%
        {selwyn2003apart}
\bibfield{author}{\bibinfo{person}{Neil Selwyn}.}
  \bibinfo{year}{2003}\natexlab{}.
\newblock \showarticletitle{Apart from technology: understanding people’s
  non-use of information and communication technologies in everyday life}.
\newblock \bibinfo{journal}{\emph{Technology in society}} \bibinfo{volume}{25},
  \bibinfo{number}{1} (\bibinfo{year}{2003}), \bibinfo{pages}{99--116}.
\newblock


\bibitem[\protect\citeauthoryear{Semuels}{Semuels}{2017}]%
        {Semuels2017Feb}
\bibfield{author}{\bibinfo{person}{Alana Semuels}.}
  \bibinfo{year}{2017}\natexlab{}.
\newblock \showarticletitle{{Why \#DeleteUber and Other Boycotts Matter }}.
\newblock \bibinfo{journal}{\emph{Atlantic}} (\bibinfo{date}{Feb}
  \bibinfo{year}{2017}).
\newblock
\urldef\tempurl%
\url{https://www.theatlantic.com/business/archive/2017/02/why-deleteuber-and-other-boycotts-matter/517416}
\showURL{%
\tempurl}


\bibitem[\protect\citeauthoryear{Sen, Ahmad, Phokeer, Farooq, Qazi, Choffnes,
  and Gummadi}{Sen et~al\mbox{.}}{2017}]%
        {sen2017inside}
\bibfield{author}{\bibinfo{person}{Rijurekha Sen}, \bibinfo{person}{Sohaib
  Ahmad}, \bibinfo{person}{Amreesh Phokeer}, \bibinfo{person}{Zaid~Ahmed
  Farooq}, \bibinfo{person}{Ihsan~Ayyub Qazi}, \bibinfo{person}{David
  Choffnes}, {and} \bibinfo{person}{Krishna~P Gummadi}.}
  \bibinfo{year}{2017}\natexlab{}.
\newblock \showarticletitle{Inside the walled garden: Deconstructing facebook's
  free basics program}.
\newblock \bibinfo{journal}{\emph{ACM SIGCOMM Computer Communication Review}}
  \bibinfo{volume}{47}, \bibinfo{number}{5} (\bibinfo{year}{2017}),
  \bibinfo{pages}{12--24}.
\newblock


\bibitem[\protect\citeauthoryear{Shafahi, Huang, Najibi, Suciu, Studer,
  Dumitras, and Goldstein}{Shafahi et~al\mbox{.}}{2018}]%
        {shafahi2018poison}
\bibfield{author}{\bibinfo{person}{Ali Shafahi}, \bibinfo{person}{W~Ronny
  Huang}, \bibinfo{person}{Mahyar Najibi}, \bibinfo{person}{Octavian Suciu},
  \bibinfo{person}{Christoph Studer}, \bibinfo{person}{Tudor Dumitras}, {and}
  \bibinfo{person}{Tom Goldstein}.} \bibinfo{year}{2018}\natexlab{}.
\newblock \showarticletitle{Poison frogs! targeted clean-label poisoning
  attacks on neural networks}. In \bibinfo{booktitle}{\emph{Advances in Neural
  Information Processing Systems}}. \bibinfo{pages}{6103--6113}.
\newblock


\bibitem[\protect\citeauthoryear{Shan, Wenger, Zhang, Li, Zheng, and Zhao}{Shan
  et~al\mbox{.}}{2020}]%
        {shan2020fawkes}
\bibfield{author}{\bibinfo{person}{Shawn Shan}, \bibinfo{person}{Emily Wenger},
  \bibinfo{person}{Jiayun Zhang}, \bibinfo{person}{Huiying Li},
  \bibinfo{person}{Haitao Zheng}, {and} \bibinfo{person}{Ben~Y. Zhao}.}
  \bibinfo{year}{2020}\natexlab{}.
\newblock \showarticletitle{Fawkes: Protecting Privacy against Unauthorized
  Deep Learning Models}. In \bibinfo{booktitle}{\emph{29th $\{$USENIX$\}$
  Security Symposium ($\{$USENIX$\}$ Security 20)}}.
  \bibinfo{pages}{1589--1604}.
\newblock


\bibitem[\protect\citeauthoryear{Silva, Santos~de Oliveira, Andreou, Vaz~de
  Melo, Goga, and Benevenuto}{Silva et~al\mbox{.}}{2020}]%
        {silva2020facebook}
\bibfield{author}{\bibinfo{person}{M{\'a}rcio Silva}, \bibinfo{person}{Lucas
  Santos~de Oliveira}, \bibinfo{person}{Athanasios Andreou},
  \bibinfo{person}{Pedro~Olmo Vaz~de Melo}, \bibinfo{person}{Oana Goga}, {and}
  \bibinfo{person}{Fabr{\'\i}cio Benevenuto}.} \bibinfo{year}{2020}\natexlab{}.
\newblock \showarticletitle{Facebook Ads Monitor: An Independent Auditing
  System for Political Ads on Facebook}. In
  \bibinfo{booktitle}{\emph{Proceedings of The Web Conference 2020}}.
  \bibinfo{pages}{224--234}.
\newblock


\bibitem[\protect\citeauthoryear{Steinhardt, Koh, and Liang}{Steinhardt
  et~al\mbox{.}}{2017}]%
        {steinhardt2017certified}
\bibfield{author}{\bibinfo{person}{Jacob Steinhardt}, \bibinfo{person}{Pang
  Wei~W Koh}, {and} \bibinfo{person}{Percy~S Liang}.}
  \bibinfo{year}{2017}\natexlab{}.
\newblock \showarticletitle{Certified defenses for data poisoning attacks}. In
  \bibinfo{booktitle}{\emph{Advances in neural information processing
  systems}}. \bibinfo{pages}{3517--3529}.
\newblock


\bibitem[\protect\citeauthoryear{Stieger, Burger, Bohn, and Voracek}{Stieger
  et~al\mbox{.}}{2013}]%
        {stieger_who_2013}
\bibfield{author}{\bibinfo{person}{Stefan Stieger}, \bibinfo{person}{Christoph
  Burger}, \bibinfo{person}{Manuel Bohn}, {and} \bibinfo{person}{Martin
  Voracek}.} \bibinfo{year}{2013}\natexlab{}.
\newblock \showarticletitle{Who commits virtual identity suicide? {Differences}
  in privacy concerns, internet addiction, and personality between {Facebook}
  users and quitters}.
\newblock \bibinfo{journal}{\emph{Cyberpsychology, Behavior, and Social
  Networking}} \bibinfo{volume}{16}, \bibinfo{number}{9}
  (\bibinfo{year}{2013}), \bibinfo{pages}{629--634}.
\newblock
\newblock
\shownote{Publisher: Mary Ann Liebert, Inc. 140 Huguenot Street, 3rd Floor New
  Rochelle, NY 10801 USA.}


\bibitem[\protect\citeauthoryear{Tahmasebian, Xiong, Sotoodeh, and
  Sunderam}{Tahmasebian et~al\mbox{.}}{2020}]%
        {tahmasebian2020crowdsourcing}
\bibfield{author}{\bibinfo{person}{Farnaz Tahmasebian}, \bibinfo{person}{Li
  Xiong}, \bibinfo{person}{Mani Sotoodeh}, {and} \bibinfo{person}{Vaidy
  Sunderam}.} \bibinfo{year}{2020}\natexlab{}.
\newblock \showarticletitle{Crowdsourcing under data poisoning attacks: A
  comparative study}. In \bibinfo{booktitle}{\emph{IFIP Annual Conference on
  Data and Applications Security and Privacy}}. Springer,
  \bibinfo{pages}{310--332}.
\newblock


\bibitem[\protect\citeauthoryear{Toma and Hancock}{Toma and Hancock}{2010}]%
        {toma2010reading}
\bibfield{author}{\bibinfo{person}{Catalina~L Toma} {and}
  \bibinfo{person}{Jeffrey~T Hancock}.} \bibinfo{year}{2010}\natexlab{}.
\newblock \showarticletitle{Reading between the lines: linguistic cues to
  deception in online dating profiles}. In
  \bibinfo{booktitle}{\emph{Proceedings of the 2010 ACM conference on Computer
  supported cooperative work}}. \bibinfo{pages}{5--8}.
\newblock


\bibitem[\protect\citeauthoryear{Troncoso}{Troncoso}{2019}]%
        {troncoso_keynote_2019}
\bibfield{author}{\bibinfo{person}{Carmela Troncoso}.}
  \bibinfo{year}{2019}\natexlab{}.
\newblock \showarticletitle{Keynote {Address}: {PETs}, {POTs}, and {Pitfalls}:
  {Rethinking} the {Protection} of {Users} against {Machine} {Learning}}.
  \bibinfo{publisher}{USENIX Association}, \bibinfo{address}{Santa Clara, CA}.
\newblock


\bibitem[\protect\citeauthoryear{Van~Kleek, Murray-Rust, Guy, O'Hara, and
  Shadbolt}{Van~Kleek et~al\mbox{.}}{2016}]%
        {10.1145/2858036.2858060}
\bibfield{author}{\bibinfo{person}{Max Van~Kleek}, \bibinfo{person}{Dave
  Murray-Rust}, \bibinfo{person}{Amy Guy}, \bibinfo{person}{Kieron O'Hara},
  {and} \bibinfo{person}{Nigel Shadbolt}.} \bibinfo{year}{2016}\natexlab{}.
\newblock \showarticletitle{Computationally Mediated Pro-Social Deception}. In
  \bibinfo{booktitle}{\emph{Proceedings of the 2016 CHI Conference on Human
  Factors in Computing Systems}} \emph{(\bibinfo{series}{CHI '16})}.
  \bibinfo{publisher}{Association for Computing Machinery},
  \bibinfo{address}{New York, NY, USA}, \bibinfo{pages}{552–563}.
\newblock
\showISBNx{9781450333627}
\urldef\tempurl%
\url{https://doi.org/10.1145/2858036.2858060}
\showDOI{\tempurl}


\bibitem[\protect\citeauthoryear{Vincent and Hecht}{Vincent and Hecht}{2021a}]%
        {vincent_can_2020}
\bibfield{author}{\bibinfo{person}{Nicholas Vincent} {and}
  \bibinfo{person}{Brent Hecht}.} \bibinfo{year}{2021}\natexlab{a}.
\newblock \showarticletitle{Can “{Conscious} {Data} {Contribution}” {Help}
  {Users} to {Exert} “{Data} {Leverage}” {Against} {Technology}
  {Companies}?}
\newblock \bibinfo{journal}{\emph{Proceedings of the ACM on Human-Computer
  Interaction}} \bibinfo{number}{CSCW}.
\newblock


\bibitem[\protect\citeauthoryear{Vincent and Hecht}{Vincent and Hecht}{2021b}]%
        {vincent2020deeper}
\bibfield{author}{\bibinfo{person}{Nicholas Vincent} {and}
  \bibinfo{person}{Brent Hecht}.} \bibinfo{year}{2021}\natexlab{b}.
\newblock \showarticletitle{A Deeper Investigation of the Importance of
  Wikipedia Links to the Success of Search Engines}.
\newblock \bibinfo{journal}{\emph{Proceedings of the ACM on Human-Computer
  Interaction}} \bibinfo{number}{CSCW} (\bibinfo{year}{2021}).
\newblock


\bibitem[\protect\citeauthoryear{Vincent, Hecht, and Sen}{Vincent
  et~al\mbox{.}}{2019a}]%
        {vincent_data_2019}
\bibfield{author}{\bibinfo{person}{Nicholas Vincent}, \bibinfo{person}{Brent
  Hecht}, {and} \bibinfo{person}{Shilad Sen}.}
  \bibinfo{year}{2019}\natexlab{a}.
\newblock \showarticletitle{“{Data} {Strikes}”: {Evaluating} the
  {Effectiveness} of {New} {Forms} of {Collective} {Action} {Against}
  {Technology} {Platforms}}. In \bibinfo{booktitle}{\emph{Proceedings of {The}
  {Web} {Conference} 2019}}.
\newblock


\bibitem[\protect\citeauthoryear{Vincent, Johnson, and Hecht}{Vincent
  et~al\mbox{.}}{2018}]%
        {vincent2018examining}
\bibfield{author}{\bibinfo{person}{Nicholas Vincent}, \bibinfo{person}{Isaac
  Johnson}, {and} \bibinfo{person}{Brent Hecht}.}
  \bibinfo{year}{2018}\natexlab{}.
\newblock \showarticletitle{Examining Wikipedia with a broader lens:
  Quantifying the value of Wikipedia's relationships with other large-scale
  online communities}. In \bibinfo{booktitle}{\emph{Proceedings of the 2018 CHI
  Conference on Human Factors in Computing Systems}}. \bibinfo{pages}{1--13}.
\newblock


\bibitem[\protect\citeauthoryear{Vincent, Johnson, Sheehan, and Hecht}{Vincent
  et~al\mbox{.}}{2019b}]%
        {vincent2019measuring}
\bibfield{author}{\bibinfo{person}{Nicholas Vincent}, \bibinfo{person}{Isaac
  Johnson}, \bibinfo{person}{Patrick Sheehan}, {and} \bibinfo{person}{Brent
  Hecht}.} \bibinfo{year}{2019}\natexlab{b}.
\newblock \showarticletitle{Measuring the importance of user-generated content
  to search engines}. In \bibinfo{booktitle}{\emph{Proceedings of the
  International AAAI Conference on Web and Social Media}},
  Vol.~\bibinfo{volume}{13}. \bibinfo{pages}{505--516}.
\newblock


\bibitem[\protect\citeauthoryear{Vincent, Li, Zha, and Hecht}{Vincent
  et~al\mbox{.}}{2019c}]%
        {vincent_mapping_2019}
\bibfield{author}{\bibinfo{person}{Nicholas Vincent}, \bibinfo{person}{Yichun
  Li}, \bibinfo{person}{Renee Zha}, {and} \bibinfo{person}{Brent Hecht}.}
  \bibinfo{year}{2019}\natexlab{c}.
\newblock \showarticletitle{Mapping the {Potential} and {Pitfalls} of "{Data}
  {Dividends}" as a {Means} of {Sharing} the {Profits} of {Artificial}
  {Intelligence}}.
\newblock \bibinfo{journal}{\emph{arXiv preprint arXiv:1912.00757}}
  (\bibinfo{year}{2019}).
\newblock


\bibitem[\protect\citeauthoryear{Waddell}{Waddell}{2020}]%
        {Waddell2020Oct}
\bibfield{author}{\bibinfo{person}{Kaveh Waddell}.}
  \bibinfo{year}{2020}\natexlab{}.
\newblock \showarticletitle{{California's New Privacy Rights Are Tough to Use,
  Consumer Reports Study Finds}}.
\newblock \bibinfo{journal}{\emph{Consum. Rep.}} (\bibinfo{date}{Oct}
  \bibinfo{year}{2020}).
\newblock
\urldef\tempurl%
\url{https://www.consumerreports.org/privacy/californias-new-privacy-rights-are-tough-to-use}
\showURL{%
\tempurl}


\bibitem[\protect\citeauthoryear{Wakabayashi}{Wakabayashi}{2018}]%
        {Wakabayashi2018Jun}
\bibfield{author}{\bibinfo{person}{Daisuke Wakabayashi}.}
  \bibinfo{year}{2018}\natexlab{}.
\newblock \showarticletitle{{California Passes Sweeping Law to Protect Online
  Privacy}}.
\newblock \bibinfo{journal}{\emph{N.Y. Times}} (\bibinfo{date}{Jun}
  \bibinfo{year}{2018}).
\newblock
\showISSN{0362-4331}
\urldef\tempurl%
\url{https://www.nytimes.com/2018/06/28/technology/california-online-privacy-law.html}
\showURL{%
\tempurl}


\bibitem[\protect\citeauthoryear{Wen, Yang, Sobolev, and Estrin}{Wen
  et~al\mbox{.}}{2018}]%
        {wen2018exploring}
\bibfield{author}{\bibinfo{person}{Hongyi Wen}, \bibinfo{person}{Longqi Yang},
  \bibinfo{person}{Michael Sobolev}, {and} \bibinfo{person}{Deborah Estrin}.}
  \bibinfo{year}{2018}\natexlab{}.
\newblock \showarticletitle{Exploring recommendations under user-controlled
  data filtering}. In \bibinfo{booktitle}{\emph{Proceedings of the 12th ACM
  Conference on Recommender Systems}}. \bibinfo{pages}{72--76}.
\newblock


\bibitem[\protect\citeauthoryear{Wilmhoff}{Wilmhoff}{2017}]%
        {wilmhoff_tom_2017}
\bibfield{author}{\bibinfo{person}{John Wilmhoff}.}
  \bibinfo{year}{2017}\natexlab{}.
\newblock \bibinfo{title}{Tom {Brady} literally owns the {Jets}, says {Google}
  search}.
\newblock
\newblock
\urldef\tempurl%
\url{https://www.espn.com/sportsnation/story/_/page/170727QTP_BradyOwnsJets/google-glitch-causes-tom-brady-appear-new-york-jets-owner%7D}
\showURL{%
\tempurl}
\newblock
\shownote{Publication Title: ESPN.}


\bibitem[\protect\citeauthoryear{Wilson and Seminario}{Wilson and
  Seminario}{2013}]%
        {wilson_when_2013}
\bibfield{author}{\bibinfo{person}{David~C Wilson} {and}
  \bibinfo{person}{Carlos~E Seminario}.} \bibinfo{year}{2013}\natexlab{}.
\newblock \showarticletitle{When power users attack: assessing impacts in
  collaborative recommender systems}. In \bibinfo{booktitle}{\emph{Proceedings
  of the 7th {ACM} conference on {Recommender} systems}}.
  \bibinfo{pages}{427--430}.
\newblock


\bibitem[\protect\citeauthoryear{Wilson and Seminario}{Wilson and
  Seminario}{2014}]%
        {wilson_evil_2014}
\bibfield{author}{\bibinfo{person}{David~C Wilson} {and}
  \bibinfo{person}{Carlos~E Seminario}.} \bibinfo{year}{2014}\natexlab{}.
\newblock \showarticletitle{Evil twins: {Modeling} power users in attacks on
  recommender systems}. In \bibinfo{booktitle}{\emph{International {Conference}
  on {User} {Modeling}, {Adaptation}, and {Personalization}}}.
  \bibinfo{publisher}{Springer}, \bibinfo{pages}{231--242}.
\newblock


\bibitem[\protect\citeauthoryear{Wyatt}{Wyatt}{2003}]%
        {wyatt2003non}
\bibfield{author}{\bibinfo{person}{Sally~ME Wyatt}.}
  \bibinfo{year}{2003}\natexlab{}.
\newblock \showarticletitle{Non-users also matter: The construction of users
  and non-users of the Internet}.
\newblock \bibinfo{journal}{\emph{Now users matter: The co-construction of
  users and technology}} (\bibinfo{year}{2003}), \bibinfo{pages}{67--79}.
\newblock


\bibitem[\protect\citeauthoryear{Xu and Zhang}{Xu and Zhang}{2013}]%
        {Xu2013Dec}
\bibfield{author}{\bibinfo{person}{Sean~Xin Xu} {and}
  \bibinfo{person}{Xiaoquan~(Michael) Zhang}.} \bibinfo{year}{2013}\natexlab{}.
\newblock \showarticletitle{{Impact of Wikipedia on Market Information
  Environment: Evidence on Management Disclosure and Investor Reaction}}.
\newblock \bibinfo{journal}{\emph{MIS Quarterly}} \bibinfo{volume}{37},
  \bibinfo{number}{4} (\bibinfo{date}{Dec} \bibinfo{year}{2013}),
  \bibinfo{pages}{1043--1068}.
\newblock
\showISSN{0276-7783}
\urldef\tempurl%
\url{http://www.jstor.org/stable/43825781}
\showURL{%
\tempurl}


\bibitem[\protect\citeauthoryear{Zhang, Monroy-Hern{\'a}ndez, Shaw, Munson,
  Gerber, Hill, Kinnaird, Farnham, and Minder}{Zhang et~al\mbox{.}}{2014}]%
        {zhang2014wedo}
\bibfield{author}{\bibinfo{person}{Haoqi Zhang}, \bibinfo{person}{Andr{\'e}s
  Monroy-Hern{\'a}ndez}, \bibinfo{person}{Aaron Shaw}, \bibinfo{person}{Sean~A
  Munson}, \bibinfo{person}{Elizabeth Gerber}, \bibinfo{person}{Benjamin~Mako
  Hill}, \bibinfo{person}{Peter Kinnaird}, \bibinfo{person}{Shelly~D Farnham},
  {and} \bibinfo{person}{Patrick Minder}.} \bibinfo{year}{2014}\natexlab{}.
\newblock \showarticletitle{WeDo: end-to-end computer supported collective
  action}. In \bibinfo{booktitle}{\emph{Eighth International AAAI Conference on
  Weblogs and Social Media}}.
\newblock


\bibitem[\protect\citeauthoryear{Zhou, Khemmarat, and Gao}{Zhou
  et~al\mbox{.}}{2010}]%
        {zhou_impact_2010}
\bibfield{author}{\bibinfo{person}{Renjie Zhou}, \bibinfo{person}{Samamon
  Khemmarat}, {and} \bibinfo{person}{Lixin Gao}.}
  \bibinfo{year}{2010}\natexlab{}.
\newblock \showarticletitle{The impact of {YouTube} recommendation system on
  video views}. In \bibinfo{booktitle}{\emph{Proceedings of the 10th {ACM}
  {SIGCOMM} conference on {Internet} measurement}}. \bibinfo{publisher}{ACM},
  \bibinfo{pages}{404--410}.
\newblock


\end{thebibliography}
